\documentclass{aa}
\usepackage{graphicx,epsfig}
\usepackage{txfonts}

\begin{document}

\title{Search for Gamma-Ray Burst Classes with the RHESSI Satellite}
\titlerunning{Search for GRB classes with RHESSI}

\author{
J. \v{R}\'{\i}pa \inst{1}
\and
A. M\'esz\'aros \inst{1}
\and
C. Wigger \inst{2,4}
\and
D. Huja \inst{1}
\and
R. Hudec \inst{3}
\and
W. Hajdas \inst{2}
}

\offprints{J. \v{R}\'{\i}pa}

\institute{Charles University,
Faculty of Mathematics and Physics,
Astronomical Institute,
V Hole\v{s}ovi\v{c}k\'ach 2, CZ 180 00 Prague 8,
Czech Republic\\
              \email{ripa@sirrah.troja.mff.cuni.cz}\\
              \email{meszaros@cesnet.cz}\\
              \email{David.HUJA@seznam.cz}
\and
Paul Scherrer Institute, CH-5232 Villigen, Switzerland\\
              \email{wojtek.hajdas@psi.ch}\\
              \email{claudia.wigger@kanti-wohlen.ch}
\and
Astronomical Institute, Academy of Sciences of the Czech
                Republic, CZ 251 65 Ond\v{r}ejov, Czech Republic\\
               \email{rhudec@asu.cas.cz}
\and
Kantonsschule Wohlen, 5610 Wohlen, Switzerland \\
}
  \date{Received September 04, 2008; accepted February 02, 2009}

\abstract
{}
{A sample of 427 gamma-ray bursts (GRBs), measured by
the RHESSI satellite, is studied statistically with respect to
duration and hardness ratio.}
{Standard statistical tests are used, such as $\chi^2$, F-test and the
maximum likelihood ratio test,
in order to compare the number of GRB groups in the RHESSI database with
that of the BATSE database.}
{Previous studies based on the BATSE Catalog claim the existence of
an intermediate GRB group, besides the long and short groups.
Using only the GRB duration $T_{90}$ as information and $\chi^2$ or F-test,
we have not found any statistically significant intermediate group in the RHESSI data.
However, maximum likelihood ratio test reveals a significant intermediate group.
Also using the 2-dimensional hardness / $T_{90}$ plane,
the maximum likelihood analysis reveals a significant
intermediate group. Contrary to the BATSE database, the
intermediate group in the RHESSI data-set is harder than the long one.}
{The existence of an intermediate group follows not only from the
BATSE data-set, but also from the RHESSI one.}


\maketitle

\section{Introduction}

In the years $1991 - 2000$ cca 3000 gamma-ray bursts (GRBs) were
detected by the BATSE instrument on board the Compton Gamma-Ray
Observatory (\cite{mee01}). After the end of this mission (June
2000) the number of discovered GRBs decreased
(down to $\sim 100-200$ GRBs annually) due to different
observational methods on the operating satellites.
Any observational GRB database from the period 2000 and latter,
can have a great importance. Such important GRB observations are
records obtained by the RHESSI satellite (\cite{rhe}) $\sim$ 70/year.

Originally it was found (\cite{kou93})
that there exist two GRB classes:
the short one with durations $\lesssim 2$\,s and the long one with
durations $\gtrsim 2$\,s.
This was confirmed with GRB data from the Konus-Wind instrument (\cite{apt98}).
However, some articles point to the existence of
three classes of GRBs in the BATSE database with respect to their durations
(\cite{ho98}, \cite{ho02}).
The work Horv\'ath et al. (2008),
using maximum likelihood ratio test on durations, gives that
there is a statistically significant intermediate group in the Swift data-set.
Also Horv\'ath et al. (2004) and Horv\'ath et al. (2006)
claimed that, when using a 2-dimensional plane of hardness ratio vs. duration,
three classes of GRBs can be found in the BATSE data-set.
\cite{mu98} pointed to the existence of three GRB
classes in multiparameter space.
In another multidimensional analysis of the BATSE catalog by \cite{cha07}
it is argued that at least three clusters of GRBs are found.
Some articles also say that the
third class (with intermediate duration), observed by BATSE,
is a bias caused by an instrumental effect (\cite{hak00}). In \cite{hak04}
there is a review and discussion of GRB classifying, based on statistical
clustering and data mining techniques, placing the intermediate group
as a separate source population in doubt.

The purpose of this paper is to investigate the number of GRB groups in
another data set, namely in the GRB data set provided by the RHESSI
satellite. Although the main goal of the RHESSI satellite
is the study of solar physics, it has a useful set of GRB observations
also covering the period $2002 - 2004$. Hence its study can be
maximally useful. Trivially, any comparison of different catalogs using different
instruments is useful for an independent confirmation
of previous results.

In the first step the 1-dimensional duration distribution
of GRBs observed by RHESSI is analysed, and in the second step
the two-dimensional plane of hardness ratio vs. duration is used.
In order to determine the number of GRB groups,
standard statistical tests described in \cite{TW53}, \cite{Press92},
\cite{NIST} are used.

The paper is organized as follows: in section \ref{sec:instrument},
the RHESSI satellite and the analysed data-set are described.
In section \ref{sec:1dim}, we present the duration distribution for the
RHESSI GRBs and its analysis. In section \ref{sec:2dim}, the 2-dimensional
hardness ratio vs. duration distribution and the maximum likelihood fit
of these data are presented.
In sections \ref{sec:discussion} and \ref{sec:conclusion},
the discussion and conclusion follow.
At the end, the RHESSI data-sample is listed.

\section{The RHESSI Data Sample}
\label{sec:instrument}

The Ramaty High Energy Solar Spectroscopic Imager (RHESSI) is a NASA Small
Explorer satellite designed to study hard X-rays and gamma-rays from solar
flares (\cite{lin02}). It consists mainly of an imaging tube and a
spectrometer. The spectrometer consists of nine germanium detectors
($7.1$\,cm in diameter and a height of 8.5\,cm) (\cite{smi02}). They are
only lightly shielded, thus making RHESSI also very useful to detect
non-solar photons from any direction (\cite{smi03}). The energy range for
GRB detection extends from about 30\,keV up to 17\,MeV.
Over a wide range of energies and GRB incoming directions,
the effective area is around 150\,cm$^2$ (\cite{claudia06}).
With a field of view of about half of the sky,
RHESSI observes about one or two GRBs per week.
Photon hits are stored event-by-event in onboard memory
with a time sampling
of $\Delta t$~=~1\,$\mu$s resolution.
The energy resolution for lines is excellent:
$\Delta E$~=~3\,keV at 1000\,keV.

We used the RHESSI GRB Catalog (\cite{rhes-grb}) and the Cosmic Burst List
(\cite{KH-list}) to find 487 GRBs in the RHESSI data between the
14$^{th}$ February 2002 and 25$^{th}$ April 2008.
We should mention the strategy how RHESSI GRBs are found.
There is no automatic GRB search routine. Only if there is
a message from any other instrument of the IPN (\cite{KH-IPN}),
the RHESSI data are searched for a GRB signal.
Therefore, in our data-set there are only GRBs,
which are also observed by other instruments.
The biggest overlap is with Konus-W. About 85\,\% of all
RHESSI GRBs are also observed by Konus-W (\cite{claudia06-talk}).

For a deeper analysis we have chosen a subset of 427 GRBs with a signal/noise ratio
better than 6. We have used the SolarSoftWare (\cite{ssw}) program running
under the Interactive Data Language (\cite{idl}) programming application as
well as our own IDL routines to derive count light-curves
(with the time resolution better than 10\,\% of the burst's duration
for the great majority of our whole data-set) and count fluences
from the rear detectors' segments (except number R2) of the
RHESSI spectrometer (\cite{smi02})
in the energy band from 25\,keV to 1.5\,MeV.
This data-set
(together with the time resolutions of derived light-curves) are
listed in the Table~\ref{data-1st} $-$ Table~\ref{data-last}.

\section{Duration Distribution}
\label{sec:1dim}

First we study the 1-dimensional duration distribution.
We use $T_{90}$ as the GRB duration, i.e. the time
interval during which the cumulative counts increase
from 5\,\% to 95\,\% above background (\cite{mee01}).
The $T_{90}$ uncertainty consists of two components.
We make an assumption that one is given by the count fluence
uncertainty during $T_{90}$ ($\delta t_{s}$),
which is given by Poisson noise, and the second one is the
time resolution of derived light-curves
($\delta t_{res}$). The total $T_{90}$ uncertainty $\delta t$
was calculated as $\delta t = \sqrt{ \delta t_s^2 + \delta_{res}^2 } . $

\begin{figure}
\centering
  \includegraphics[width=91mm]{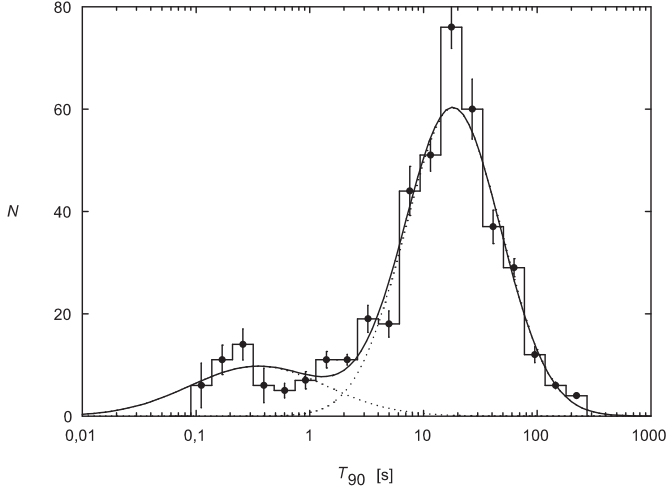}
  \caption{Duration distribution of the 427 RHESSI
           bursts with the best $\chi^2$ fit of two log-normal functions.
           Number of bins is 19, $dof = 14$ and
           $\chi^2 \simeq 19.1$ which implies the goodness-of-fit $\simeq 16\,\%$.
           The bar errors are standard deviations of the number of GRBs
           per bin for ten different simulated duration distributions as
           described in the text.}
  \label{2gauss}
\end{figure}

The histogram of the times $T_{90}$ gives a distribution with
two maxima: at approximately 0.2\,s and 20\,s (Fig.~\ref{2gauss}.).
The histogram consists of 19 equally wide bins
on logarithmic scale (with base 10)
starting at 0.09\,s and ending at 273.4\,s.

\begin{figure}
\centering
  \includegraphics[width=91mm]{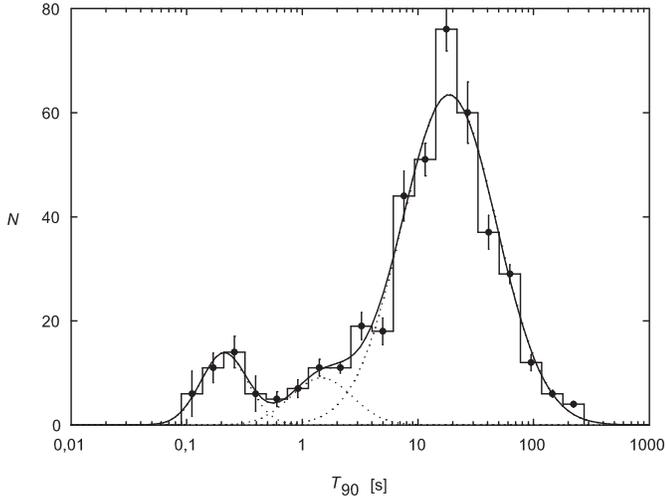}
  \caption{Duration distribution of the 427 RHESSI bursts
with the best $\chi^2$ fit of three log-normal functions.
           Number of bins is 19, $dof = 11$
and $\chi^2 \simeq 10.3$ which implies the goodness-of-fit $\simeq 50\,\%$.
The bar errors are the same as described in Fig.~\ref{2gauss}.}
  \label{3gauss}
\end{figure}

We follow the method done by Horv\'ath (1998)
and fitted one, two (Fig.~\ref{2gauss}.) and three (Fig.~\ref{3gauss}.)
log-normal functions and used the $\chi^2$ test to
evaluate these fits.
The minimal number of GRBs per bin is 4 (last bin),
hence the use of the $\chi^2$ test is possible.

In the case of the fit with one log-normal function we
obtained $\chi^2 \simeq 157$ for 17 degrees of freedom (dof).
Therefore, this hypothesis is rejected
on a smaller than 0.01\,\% significance level.

The fit with two log-normal functions
is shown in  Fig.~\ref{2gauss}  and
the  fit with three log-normal functions  in Fig.~\ref{3gauss}.
The parameters of the fits, the values of $\chi^2$,
the degrees of freedom
and the goodness-of-fits are listed in Table~\ref{parametry-T90}.

The assumption of two groups being represented by two
log-normal fits is acceptable,
the fit with three log-normal functions
even more.
The question is whether the improvement in
$\chi^2$ is statistically significant.
To answer this question, we used the F-test,
as described by \cite{ba97}.
The F-test gives a probability of 6.9\,\%
of the improvement in $\chi^2$ being accidental.
This value is remarkably low, but not
low enough to reject the hypothesis that two log-normal
functions are still enough to describe the observed duration distribution.

In order to know how the $T_{90}$ uncertainties effect our result,
we randomly picked up one half of the bursts and shifted their durations
by the full amount of their uncertainties to lower values and the second half
to higher values. Then we made a histogram and recalculated the
best fitted parameters, $\chi^2$ and F-test.
The results for ten such calculations are listed in the
Table~\ref{parametry-T90-chyby}. This method also gives us information
how the fitted parameters vary, and thus tells what are their uncertainties.
From Table~\ref{parametry-T90-chyby} we see that, on average, the improvement in
$\chi^2$ is not significant. Therefore, we can not
proclaim acceptance of three groups by using this statistical method.

Since the number of GRBs is low for many bins,
we also used the maximum likelihood method
(see \cite{ho02} and the references therein)
in order to fit two and three log-normal functions
on the RHESSI data-set.

The parameters are listed in the Table~\ref{parametry-T90-ML-method}.

As the difference of the
logarithms of the likelihoods $\Delta \ln L = 9.2$ should be half of
the $\chi^2$ distribution for 3 degrees of freedom (\cite{ho02}), we
obtain that the introduction of a third group is statistically
significant on the 0.036\,\% level (of being accidental).

To get an image how $T_{90}$ uncertainties effect our result,
we proceed similarly as in the $\chi^2$ fitting and generated ten
different data-sets randomly changed in durations by the full amount of their uncertainties.
The results are presented in the Table~\ref{parametry-T90-ML-method-chyby}.
From this table it is seen that all ten simulations give probabilities,
that introducing of the third group is accidental, much lower than 5\,\%.
Thus, the hypothesis of introducing third group is highly acceptable.

\begin{center}
\begin{table} \caption{Parameters of the best $\chi^2$ fits of
two and three log-normal functions on the RHESSI GRB $T_{90}$ distribution.
$\mu$ are the means, $\sigma$ are the standard deviations
and $w$ are the weights of the distribution.
Given uncertainties are standard deviations of the parameters
obtained by ten different fittings of randomly changed histogram of durations
by their uncertainties.}
\centering
\begin{tabular}{lrr}

\hline
\\[-2ex]
parameter                    &    2 log-normal  &  3 log-normal  \\[0.3ex]
\hline\hline
\\[-2ex]
$\mu_{\textrm{short}}$       &     -0.46$\pm$0.13   &  -0.68$\pm$0.09  \\[0.3ex]
$\sigma_{\textrm{short}}$    &      0.60$\pm$0.06   &   0.20$\pm$0.03  \\[0.3ex]
$w_{\textrm{short}}[\%]$     &     18.7$\pm$1.5     &   8.9$\pm$1.0    \\[0.3ex]
\\[-2ex]
\hline
$\mu_{\textrm{long}}$        &      1.26$\pm$0.03   &  1.27$\pm$0.03  \\[0.3ex]
$\sigma_{\textrm{long}}$     &      0.42$\pm$0.01   &  0.41$\pm$0.01  \\[0.3ex]
$w_{\textrm{long}}[\%]$      &      81.3$\pm$1.5    &  83.4$\pm$0.9    \\[0.3ex]
\\[-2ex]
\hline
$\mu_{\textrm{middle}}$      &                  &  0.17$\pm$0.06 \\[0.3ex]
$\sigma_{\textrm{middle}}$   &                  &  0.27$\pm$0.06 \\[0.3ex]
$w_{\textrm{middle}}[\%]$    &                  &   7.7$\pm$1.0   \\[0.3ex]
\hline
\\[-2ex]
$dof$                        &          14      &            11    \\[0.3ex]
$\chi^2$                     &          19.13   &            10.30 \\[0.3ex]
goodness[\%]                 &          16.0    &            50.4  \\[0.3ex]
\hline
\\[-2ex]
$ F_{0} $                    &          3.14    &                  \\[0.3ex]
$P(F>F_{0})[\%] $            &           6.9    &                  \\[0.3ex]
\\[-2ex]
\hline
\label{parametry-T90}
\end{tabular}
\end{table}
\end{center}

\begin{center}
\begin{table} \caption{The minimal $\chi^2$, corresponding goodness-of-fits
and F-tests for fitted two and three log-normal functions
on the RHESSI GRB $T_{90}$ distribution for ten different changes of durations
by their uncertainties.}
\centering
\begin{tabular}{rr|rr|r}

\hline
\\[-2ex]
               &  2 log-normal  &                 &  3 log-normal  &  F-test             \\[0.3ex]
\\[-2ex]
\hline
\\[-2ex]
$\chi^2$       &  goodness      &  $\chi^2$       &  goodness      &  $P(F>F_{0})$       \\[0.3ex]
               &    [\%]        &                 &     [\%]       &     [\%]            \\[0.3ex]
\\[-2ex]
\hline\hline
\\[-2ex]
   23.97       &     4.6        &    14.19        &    22.3        &   11.1              \\[0.3ex]
   18.03       &    20.6        &    10.65        &    47.3        &   11.0              \\[0.3ex]
   15.99       &    31.4        &     5.13        &    92.5        &    0.5              \\[0.3ex]
   13.52       &    48.6        &     7.50        &    75.8        &    8.0              \\[0.3ex]
   17.87       &    21.3        &     7.55        &    75.3        &    2.0              \\[0.3ex]
   16.36       &    29.2        &     9.66        &    56.1        &   11.0              \\[0.3ex]
   11.89       &    61.5        &     6.95        &    80.3        &   10.4              \\[0.3ex]
   21.86       &     8.2        &    13.77        &    24.6        &   15.1              \\[0.3ex]
   20.07       &    12.8        &     9.49        &    57.7        &    3.5              \\[0.3ex]
   20.40       &    11.8        &    12.73        &    31.1        &   14.4              \\[0.3ex]
\\[-2ex]
\hline
\label{parametry-T90-chyby}
\end{tabular}
\end{table}
\end{center}

\begin{center}
\begin{table} \caption{Parameters of the best fit
with two and three log-normal functions
done by the maximum likelihood method on the RHESSI data.
$\mu$ are the means, $\sigma$ are the standard deviations,
$w$ are the weights of the distribution and
$L_{\textrm{2}}, L_{\textrm{3}}$ are the likelihoods.
Given uncertainties are standard deviations of the parameters
obtained by ten different fittings of data-sets, in which the durations
were randomly changed by their uncertainties.}
\centering
\begin{tabular}{lrr}

\hline
\\[-2ex]
parameter                    & 2 log-normal        & 3 log-normal \\[0.3ex]
\hline\hline
\\[-2ex]
$\mu_{\textrm{short}}$       &    -0.60$\pm$0.07   &   -0.64$\pm$0.02 \\[0.3ex]
$\sigma_{\textrm{short}}$    &    0.25$\pm$0.05    &    0.20$\pm$0.02 \\[0.3ex]
$w_{\textrm{short}}[\%]$     &    10.2$\pm$1.3     &   9.4$\pm$0.4   \\[0.3ex]
\\[-2ex]
\hline
$\mu_{\textrm{long}}$        &    1.20$\pm$0.01    &    1.26$\pm$0.01 \\[0.3ex]
$\sigma_{\textrm{long}}$     &    0.47$\pm$0.01    &    0.41$\pm$0.01 \\[0.3ex]
$w_{\textrm{long}}[\%]$      &    89.8$\pm$1.3     &    84.4$\pm$1.0  \\[0.3ex]
\\[-2ex]
\hline
$\mu_{\textrm{middle}}$      &                  &   0.17$\pm$0.04 \\[0.3ex]
$\sigma_{\textrm{middle}}$   &                  &   0.22$\pm$0.06 \\[0.3ex]
$w_{\textrm{middle}}[\%]$    &                  &    6.2$\pm$1.4  \\[0.3ex]
\hline
\\[-2ex]
$\ln L_{\textrm{2}}$         &    -389.17   &          \\[0.3ex]
$\ln L_{\textrm{3}}$         &              & -379.95  \\[0.3ex]
\\[-2ex]
\hline
\label{parametry-T90-ML-method}
\end{tabular}
\end{table}
\end{center}

\begin{center}
\begin{table} \caption{The maximal likelihoods and
corresponding probabilities that introducing of the third group is accidental
for maximum likelihood fittings (one-dimensional) with two and three log-normal functions of
ten different changes of durations by their uncertainties.}
\centering
\begin{tabular}{ccc}

\hline
\\[-2ex]
$\ln L_{\textrm{2}}$  &  $\ln L_{\textrm{3}}$  &  probability  \\[0.3ex]
                      &                        &     [\%]      \\[0.3ex]
\\[-2ex]
\hline\hline
\\[-2ex]
     -388.12          &      -378.86           &    0.03      \\[0.3ex]
     -390.82          &      -383.92           &    0.32      \\[0.3ex]
     -391.90          &      -380.97           &    0.01      \\[0.3ex]
     -391.75          &      -385.37           &    0.52      \\[0.3ex]
     -392.25          &      -384.24           &    0.11      \\[0.3ex]
     -390.62          &      -383.67           &    0.30      \\[0.3ex]
     -386.26          &      -375.54           &    0.01      \\[0.3ex]
     -392.33          &      -384.97           &    0.21      \\[0.3ex]
     -389.16          &      -380.93           &    0.09      \\[0.3ex]
     -390.21          &      -384.32           &    0.82      \\[0.3ex]
\\[-2ex]
\hline
\label{parametry-T90-ML-method-chyby}
\end{tabular}
\end{table}
\end{center}

\section{Hardness Ratio vs. Duration}
\label{sec:2dim}

\begin{figure}[]
  \centerline{\epsfxsize=94mm \epsfbox[1 1 595 363]{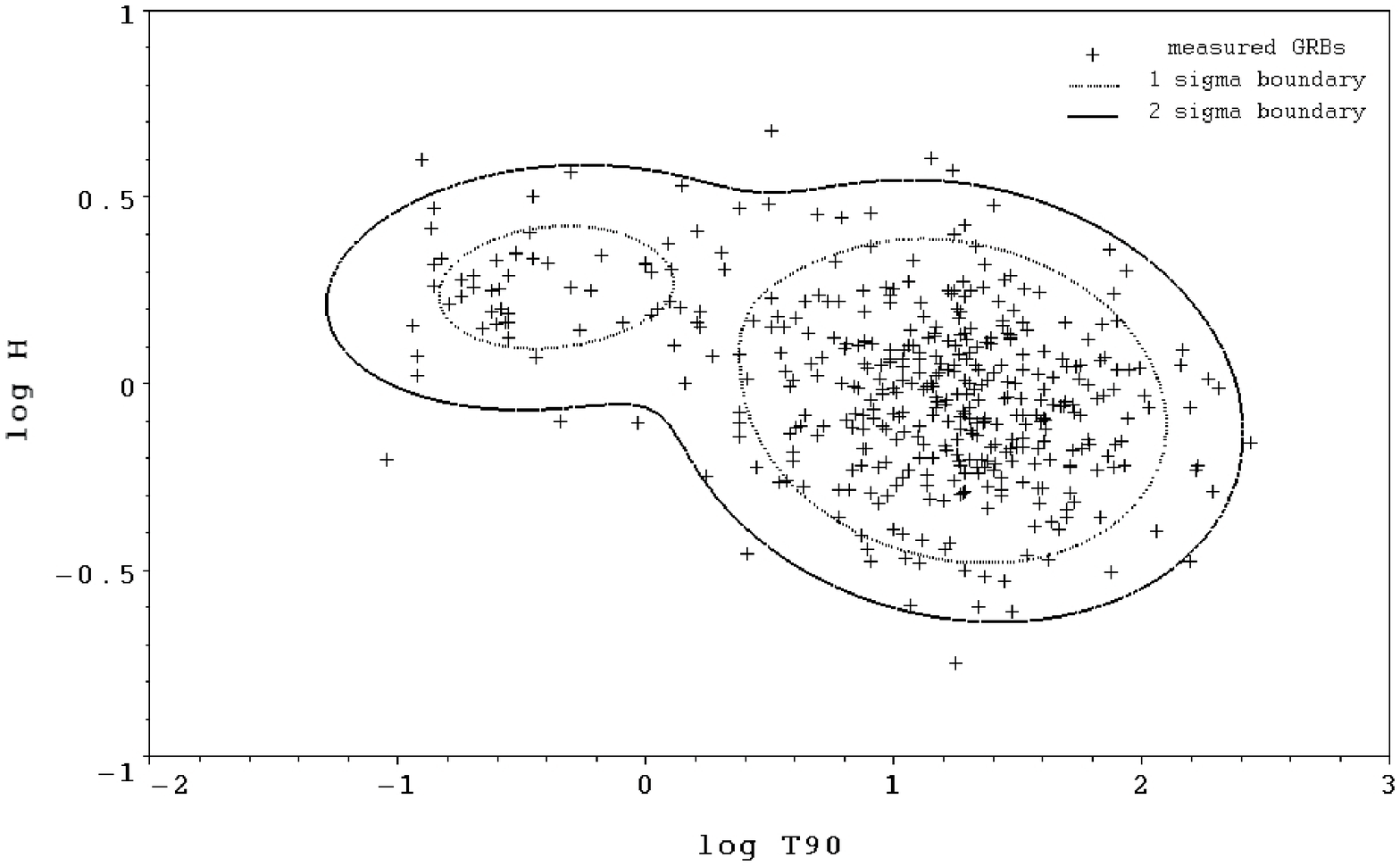}}
  \caption{Hardness ratio vs. $T_{90}$ of the RHESSI GRBs with
the best fit of two bivariate log-normal functions.}
  \label{hardness2}
\end{figure}

\begin{figure}[]
  \centerline{\epsfxsize=94mm \epsfbox[1 1 595 363]{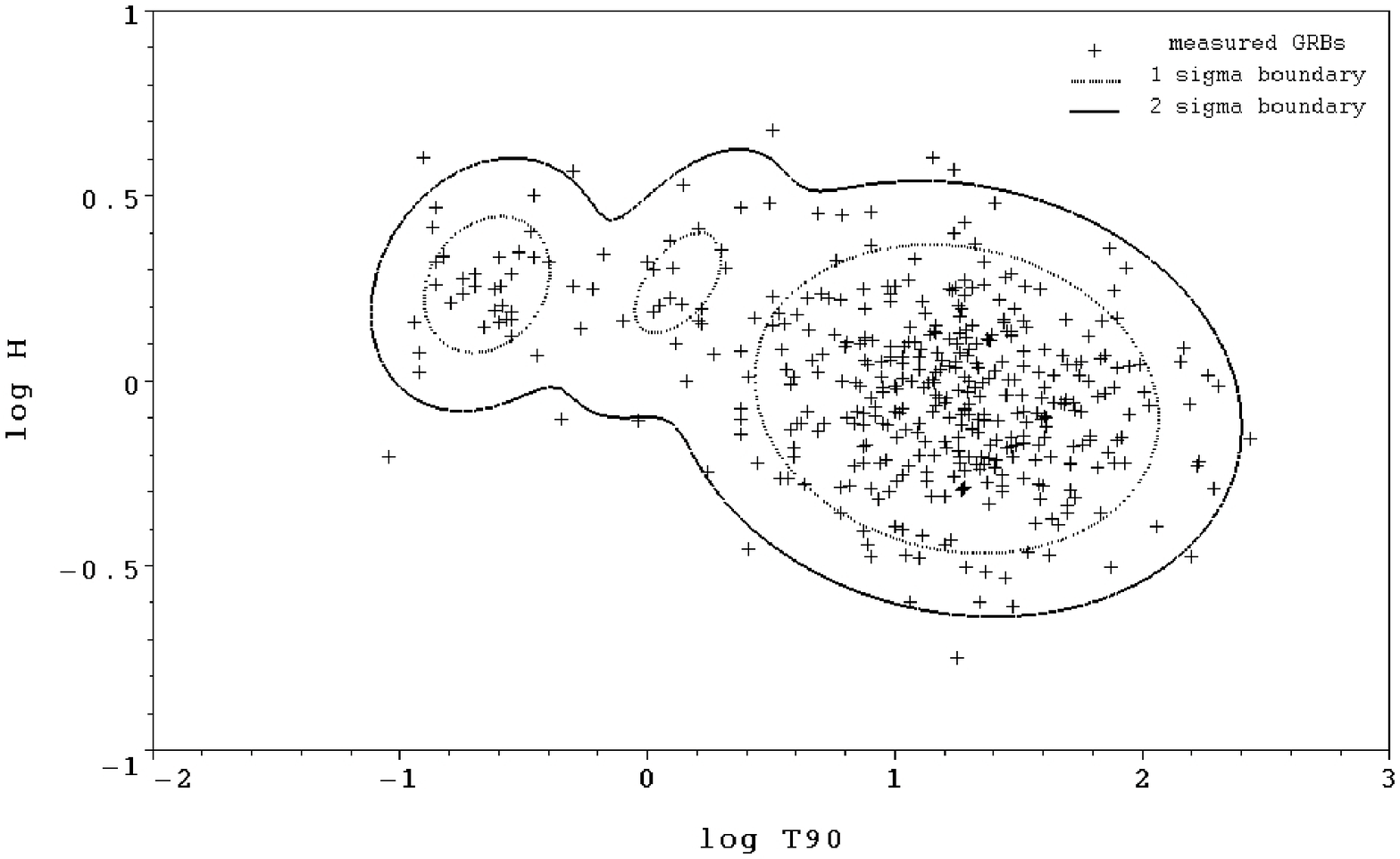}}
  \caption{Hardness ratio vs. $T_{90}$ of the RHESSI GRBs with
the best fit of three bivariate log-normal functions.}
  \label{hardness3}
\end{figure}

A two-dimensional scatter plot of RHESSI GRBs is shown in
Fig.~\ref{hardness2} and Fig.~\ref{hardness3}.
One axis is the duration $T_{90}$,
used in the previous section, the other axis
is a hardness ratio.
The hardness ratio is defined as a ratio of
two fluences $F$ in two different energy bands
integrated over the time interval $T_{90}$.
For the RHESSI data-set, we used the energy bands
$(25 - 120)$\,keV and $(120 - 1500)$\,keV, i.e. $H=F_{120-1500}/F_{25-120}$.

Using the maximum likelihood method (see Horv\'ath et al. (2004),
Horv\'ath et al. (2006) and the references therein),
we fit two and three bivariate log-normal functions
in order to search for clusters.
In Fig.~\ref{hardness2}., we show
the best fit of two bivariate log-normal functions
(11 independent parameters, since
the two weights must add up to 100\,\%).

The parameters are listed in Table~\ref{parametry-hardness}.
One result is that the short GRBs are
on average harder than long GRBs.
Having a closer look at the GRB distribution within the short class,
one can see that the points within the 1$\sigma$ ellipse are not
evenly distributed. They cluster towards the shortest durations
(Fig.~\ref{hardness2}).

The fitting with the sum of three groups (17 independent
parameters) is shown in Fig.~\ref{hardness3}.
The fitted parameters are listed in Table~\ref{parametry-hardness}.
The former short group is clearly separated into two parts.
As far as one can tell by sight, the data points scatter evenly
within (and around) the 1$\sigma$ ellipses.

As the difference of the
logarithms of the likelihoods $\Delta \ln L = 10.9$ should be half of
the $\chi^2$ distribution for 6 degrees of freedom (\cite{ho06}), we
obtain that the introduction of a third group is statistically
significant on the 0.13\,\% level (of being accidental).

To get an image how GRB durations and hardness ratio uncertainties effect our result,
we proceed similarly as in the $\chi^2$ fitting and generated ten
different data-sets randomly changed in durations and hardness ratios
by the full amount of their uncertainties.
The results are presented in the Table~\ref{parametry-hardness-chyby}.
From  this table it is seen that almost all simulations give probabilities,
that introducing of the third group is accidental, much lower than 5\,\%.
Thus, the hypothesis of introducing the third group is highly acceptable.

\begin{center}
\begin{table} \caption{Parameters of the best fit
with two and three bivariate log-normal functions
done by the maximum likelihood method on the RHESSI data.
$\mu_{\textrm{x}}$ are the means on the $x$-axis ($x = \log T_{90}$),
$\mu_{\textrm{y}}$ are the means on the $y$-axis ($y = \log H$),
$\sigma_{\textrm{x}}$ are the dispersions on the $x$-axis,
$\sigma_{\textrm{y}}$ are the dispersions on the $y$-axis,
$r$ are the correlation coefficients, $w$ are
the weights of the distribution and $L_{\textrm{2}}, L_{\textrm{3}}$
are the likelihoods.
Given uncertainties are standard deviations of the parameters
obtained by ten different fittings of data-sets, where the durations
and hardness ratios were randomly changed by their uncertainties.}
\centering
\begin{tabular}{lrr}

\hline
\\[-2ex]
parameter & 2 log-normal & 3 log-normal \\[0.3ex]
\hline\hline
\\[-2ex]
$\mu_{\textrm{x,short}}$     &     -0.38$\pm$0.018    & -0.65$\pm$0.018    \\[0.3ex]
$\mu_{\textrm{y,short}}$     &      0.26$\pm$0.010    &  0.26$\pm$0.009    \\[0.3ex]
$\sigma_{\textrm{x,short}}$  &      0.42$\pm$0.012    &  0.20$\pm$0.016    \\[0.3ex]
$\sigma_{\textrm{y,short}}$  &      0.15$\pm$0.009    &  0.15$\pm$0.012    \\[0.3ex]
$w_{\textrm{short}}[\%]$     &      14.2$\pm$0.3      &   9.2$\pm$0.5      \\[0.3ex]
$r_{\textrm{short}}$         &      0.14$\pm$0.092    &  0.22$\pm$0.119    \\[0.3ex]
\\[-2ex]
\hline
$\mu_{\textrm{x,long}}$      &      1.25$\pm$0.004    &  1.25$\pm$0.006    \\[0.3ex]
$\mu_{\textrm{y,long}}$      &     -0.05$\pm$0.004    & -0.05$\pm$0.004    \\[0.3ex]
$\sigma_{\textrm{x,long}}$   &      0.42$\pm$0.004    &  0.42$\pm$0.005    \\[0.3ex]
$\sigma_{\textrm{y,long}}$   &      0.22$\pm$0.003    &  0.22$\pm$0.003    \\[0.3ex]
$w_{\textrm{long}}[\%]$      &      85.8$\pm$0.3      &   85.5$\pm$0.8     \\[0.3ex]
$r_{\textrm{long}}$          &     -0.14$\pm$0.018    & -0.13$\pm$0.020    \\[0.3ex]
\\[-2ex]
\hline
$\mu_{\textrm{x,middle}}$    &              &  0.11$\pm$0.029    \\[0.3ex]
$\mu_{\textrm{y,middle}}$    &              &  0.27$\pm$0.019    \\[0.3ex]
$\sigma_{\textrm{x,middle}}$ &              &  0.21$\pm$0.057    \\[0.3ex]
$\sigma_{\textrm{y,middle}}$ &              &  0.17$\pm$0.035    \\[0.3ex]
$w_{\textrm{middle}}[\%]$    &              &   5.3$\pm$1.1      \\[0.3ex]
$r_{\textrm{middle}}$        &              &  0.59$\pm$0.230    \\[0.3ex]
\\[-2ex]
\hline
$\ln L_{\textrm{2}}$         &   -323.91    &          \\[0.3ex]
$\ln L_{\textrm{3}}$         &              & -313.00  \\[0.3ex]
\hline
\label{parametry-hardness}
\end{tabular}
\end{table}
\end{center}

\begin{center}
\begin{table} \caption{The maximal likelihoods and
corresponding probabilities that introducing of the third group is accidental
for maximum likelihood fittings (two-dimensional) with two and three
bivariate log-normal functions of ten different changes of
durations and hardness ratios by their uncertainties.}
\centering
\begin{tabular}{ccc}

\hline
\\[-2ex]
$\ln L_{\textrm{2}}$  &  $\ln L_{\textrm{3}}$  &  probability  \\[0.3ex]
                      &                        &     [\%]      \\[0.3ex]
\\[-2ex]
\hline\hline
\\[-2ex]
-349.40        &  -339.92        &  0.42    \\[0.3ex]
-348.28        &  -336.36        &  0.06    \\[0.3ex]
-350.66        &  -340.80        &  0.31    \\[0.3ex]
-347.12        &  -338.15        &  0.64    \\[0.3ex]
-349.64        &  -340.47        &  0.54    \\[0.3ex]
-344.59        &  -338.17        &  4.56    \\[0.3ex]
-350.36        &  -342.47        &  1.50    \\[0.3ex]
-344.44        &  -334.71        &  0.35    \\[0.3ex]
-349.37        &  -336.51        &  0.03    \\[0.3ex]
-348.06        &  -338.57        &  0.42    \\[0.3ex]
\\[-2ex]
\hline
\label{parametry-hardness-chyby}
\end{tabular}
\end{table}
\end{center}

\section{Discussion}
\label{sec:discussion}

The analysis of the 1-dimensional duration distribution,
by the $\chi^2$ fitting, reveals
the class of so called long GRBs (about 83\,\% of all RHESSI GRBs)
with typical durations from 5 to 70 seconds,
the most probable duration being $T_{90} \approx 19$\,s.
Another class are short GRBs (about 9\,\% of all RHESSI GRBs)
with typical durations from 0.1 to 0.4 seconds,
the most probable duration being $T_{90} \approx 0.21$\,s.
By fitting 3 log-normal functions we find
a third class (about 8\,\% of all RHESSI GRBs)
with typical durations from 0.8 to 3 seconds,
the most probable duration being $T_{90} \approx 1.5$\,s.
The existence of the intermediate class from the RHESSI
$T_{90}$ distribution is not confirmed on a sufficiently high
significance using only the $\chi^2$ fit.
However, the maximum likelihood ratio test
on the same data reveals that the introduction
of a third class is statistically significant.
The $\chi^2$ method might not be as much sensitive and hance decisive
as the likelihood method, because of the low number
of bursts in our data-sample (\cite{ho08}, $2^{nd}$ section,
$1^{st}$ paragraph).

The hardness ratio vs. duration plot for the RHESSI sample
does further demonstrate the existence of a third class.
The typical durations are very similar to the ones
obtained with the 1-dimensional analysis, the percentages
are slightly different ($\approx$ 86\,\% long,
$\approx$ 9\,\% short, $\approx$ 5\,\% intermediate).

Three classes of GRBs were also reported for the
BATSE GRBs (\cite{ho06}) and the Swift GRBs (\cite{ho08}).
For BATSE, $\approx$ 65\,\% of all GRBs are long,
$\approx$ 24\,\% short and  $\approx$ 11\,\% intermediate
(\cite{ho06} Table~2. of that article).
The typical durations found for BATSE are
roughly a factor 2 longer than for RHESSI,
but consistent for all three classes.
As is known from BATSE, also in the RHESSI data-set,
the short GRBs are on average harder than
the long GRBs.
The most remarkable difference is
the hardness of the intermediate class.
In the BATSE data, the intermediate class has the lowest
hardness ratio, which is anti-correlated with the duration (\cite{ho06}),
whereas we find for the RHESSI data that its hardness is
comparable with that of the short group  and
correlated with the duration,
but this correlation is not conclusive
because of its large error.
The hardness of the intermediate class found with RHESSI
is surprising since the intermediate class
in the BATSE data was found to be the softest.
This discrepancy might by explained by the different
definitions of the hardnesses. The hardness $H$ for the RHESSI data is defined
as $H=F_{120-1500}/F_{25-120}$, whereas for the BATSE data $H=F_{100-320}/F_{50-100}$, where the
numbers denotes energy in keV (the BATSE fluences at higher energies than 320\,keV are noisy (\cite{bag98})).
This means that hardnesses do not measure the same bursts' behaviours. Even more different is the situation if
we compare hardnesses in the Swift and RHESSI databases, because the Swifts' hardnesses are defined as
$H=F_{100-150}/F_{50-100}$ and $H=F_{50-100}/F_{25-50}$ (\cite{ho08}, \cite{sak08}).

The shorter durations of the RHESSI GRBs compared
to the BATSE GRBs can be understood:
For RHESSI, which is practically unshielded,
the background is high
(minimum around 1000 counts per second  in the $(25 - 1500)$\,keV band)
and varies by up to a factor 3.
Additionally, RHESSI's sensitivity drops rapidly below $\approx$ 50\,keV.
Weak GRBs (in the sense of counts per second) and soft GRBs
are not so well observed by RHESSI. Since GRBs tend to
be softer and weaker at later times, they will sooner
fall beyond RHESSI's detection limit, resulting
in a shorter duration.

For Swift,  $\approx$ 58\% of all GRBs are long,
$\approx$ 7\% short and  $\approx$ 35\% intermediate (\cite{ho08}).
The percentage of each group depends obviously
on the instrument.

\section{Conclusion}
\label{sec:conclusion}

The RHESSI data confirm that GRBs can be separated
into a short and long class, and that the short GRBs
are on average harder than the long ones.
The two-dimensional analysis in the hardness/duration plane
as well as the maximum likelihood fit of the duration
distribution show a third class with intermediate
duration and similar hardness as the short class.

\begin{acknowledgements}

This study was supported by the GAUK grant No. 46307,
by the OTKA grants No. T48870 and K 77795, by the Grant Agency of
the Czech Republic grant No. 205/08/H005,
by the Research Program MSM0021620860 of the Ministry of Education
of the Czech Republic, by the INTEGRAL PECS Project 98023
and by the grant GA \v{C}R 205/08/1207.
We appreciate help of K. Hurley with the RHESSI GRB list,
valuable discussion with L.G. Bal\'azs and useful remarks of O. Wigger.
Thanks are due to the anonymous referee for the worthwhile notes.

\end{acknowledgements}

\begin{center}
\begin{table} \caption{The RHESSI GRB data-set including
I. GRB names which correspond to dates (the letters after GRB names
are internal and do not have to be in accordance with e.g. GCN GRB names),
II. GRB peak time, III. $T_{90}$ duration,
IV. time resolution $\delta t_{res}$ (described above) and
V. hardness ratios.}
\centering
\label{data-1st}
\begin{tabular}{lrrrr}
\hline
\\[-2ex]
GRB     & peak time         & $T_{90}$                          & $\delta t_{res}$ & hardness ratio         \\[0.3ex]
        & UTC               & [s]                               & [s]              & log $H$                \\[0.3ex]
\hline\hline
\\[-2ex]

\tiny   020214	&\tiny 18:49:47.700	&\tiny (1.42$\pm$0.03)E+1	&\tiny 2.0E-1	&\tiny (6.03$\pm$0.16)E-1	\\
\tiny	020218	&\tiny 19:49:41.750	&\tiny (3.40$\pm$0.06)E+1	&\tiny 5.0E-1	&\tiny -(1.12$\pm$0.11)E-1	\\
\tiny	020302	&\tiny 12:23:54.400	&\tiny (4.88$\pm$0.32)E+1	&\tiny 8.0E-1	&\tiny (1.65$\pm$0.53)E-1	\\
\tiny	020306	&\tiny 18:58:02.893	&\tiny (1.35$\pm$0.16)E-1	&\tiny 1.5E-2	&\tiny (4.16$\pm$0.40)E-1	\\
\tiny	020311	&\tiny 01:21:31.550	&\tiny (1.08$\pm$0.06)E+1	&\tiny 3.0E-1	&\tiny (6.55$\pm$3.88)E-2	\\
\tiny	020313	&\tiny 01:17:53.400	&\tiny (2.32$\pm$0.10)E+1	&\tiny 4.0E-1	&\tiny -(5.17$\pm$0.50)E-1	\\
\tiny	020315	&\tiny 15:42:47.550	&\tiny (1.26$\pm$0.11)E+1	&\tiny 3.0E-1	&\tiny (2.97$\pm$6.94)E-2	\\
\tiny	020331	&\tiny 10:23:26.750	&\tiny (3.90$\pm$0.53)E+1	&\tiny 5.0E-1	&\tiny -(0.46$\pm$1.05)E-1	\\
\tiny	020407	&\tiny 04:14:44.300	&\tiny (2.10$\pm$0.09)E+1	&\tiny 6.0E-1	&\tiny -(1.72$\pm$2.76)E-2	\\
\tiny	020409	&\tiny 09:27:23.500	&\tiny (1.46$\pm$0.14)E+2	&\tiny 1.0E+0	&\tiny (8.84$\pm$7.37)E-2	\\
\tiny	020413	&\tiny 16:20:15.500	&\tiny (9.00$\pm$0.64)E+0	&\tiny 2.0E-1	&\tiny -(2.89$\pm$5.57)E-2	\\
\tiny	020417	&\tiny 05:36:26.250	&\tiny (7.85$\pm$0.52)E+1	&\tiny 5.0E-1	&\tiny -(1.62$\pm$0.58)E-1	\\
\tiny	020418	&\tiny 17:43:08.850	&\tiny (3.40$\pm$0.13)E+0	&\tiny 1.0E-1	&\tiny (1.81$\pm$0.20)E-1	\\
\tiny	020426	&\tiny 23:56:14.795	&\tiny (1.20$\pm$0.32)E-1	&\tiny 3.0E-2	&\tiny (7.40$\pm$7.13)E-2	\\
\tiny	020430	&\tiny 21:22:01.650	&\tiny (1.35$\pm$0.05)E+1	&\tiny 3.0E-1	&\tiny -(2.73$\pm$0.32)E-1	\\
\tiny	020509	&\tiny 00:01:17.675	&\tiny (3.50$\pm$0.38)E+0	&\tiny 5.0E-2	&\tiny (8.32$\pm$8.52)E-2	\\
\tiny	020524	&\tiny 02:12:47.300	&\tiny (1.48$\pm$0.08)E+1	&\tiny 2.0E-1	&\tiny (7.46$\pm$4.26)E-2	\\
\tiny	020525A	&\tiny 03:47:53.650	&\tiny (7.60$\pm$0.61)E+0	&\tiny 1.0E-1	&\tiny (1.10$\pm$0.65)E-1	\\
\tiny	020525B	&\tiny 04:26:53.210	&\tiny (1.80$\pm$0.29)E-1	&\tiny 2.0E-2	&\tiny (2.78$\pm$0.96)E-1	\\
\tiny	020527	&\tiny 05:17:20.525	&\tiny (1.65$\pm$0.18)E+0	&\tiny 5.0E-2	&\tiny (1.94$\pm$0.88)E-1	\\
\tiny	020602	&\tiny 17:30:28.085	&\tiny (1.75$\pm$0.19)E+0	&\tiny 7.0E-2	&\tiny -(2.49$\pm$0.91)E-1	\\
\tiny	020603	&\tiny 17:50:34.905	&\tiny (1.44$\pm$0.06)E+0	&\tiny 3.0E-2	&\tiny -(0.07$\pm$3.08)E-2	\\
\tiny	020604	&\tiny 14:13:42.850	&\tiny (7.20$\pm$0.41)E+0	&\tiny 3.0E-1	&\tiny -(1.37$\pm$3.19)E-2	\\
\tiny	020620	&\tiny 12:58:06.875	&\tiny (3.45$\pm$0.22)E+0	&\tiny 5.0E-2	&\tiny -(2.63$\pm$0.56)E-1	\\
\tiny	020623	&\tiny 04:23:07.350	&\tiny (6.60$\pm$0.46)E+0	&\tiny 1.0E-1	&\tiny -(2.84$\pm$0.66)E-1	\\
\tiny	020630	&\tiny 07:58:52.650	&\tiny (3.30$\pm$0.26)E+1	&\tiny 3.0E-1	&\tiny -(1.48$\pm$0.66)E-1	\\
\tiny	020702	&\tiny 15:54:16.250	&\tiny (3.69$\pm$0.24)E+1	&\tiny 3.0E-1	&\tiny -(3.85$\pm$0.71)E-1	\\
\tiny	020708	&\tiny 04:34:12.500	&\tiny (2.70$\pm$0.40)E+1	&\tiny 1.0E+0	&\tiny -(3.00$\pm$1.29)E-1	\\
\tiny	020712	&\tiny 06:09:44.100	&\tiny (1.80$\pm$0.20)E+1	&\tiny 2.0E-1	&\tiny -(1.51$\pm$0.93)E-1	\\
\tiny	020715A	&\tiny 15:14:26.735	&\tiny (2.40$\pm$0.40)E-1	&\tiny 3.0E-2	&\tiny (1.92$\pm$0.88)E-1	\\
\tiny	020715B	&\tiny 19:21:09.100	&\tiny (6.40$\pm$0.21)E+0	&\tiny 2.0E-1	&\tiny (1.05$\pm$0.10)E-1	\\
\tiny	020725	&\tiny 16:25:41.150	&\tiny (5.00$\pm$0.31)E+0	&\tiny 1.0E-1	&\tiny (2.37$\pm$0.51)E-1	\\
\tiny	020801	&\tiny 11:52:41.550	&\tiny (1.43$\pm$0.09)E+1	&\tiny 1.0E-1	&\tiny -(8.54$\pm$5.30)E-2	\\
\tiny	020819A	&\tiny 07:56:39.455	&\tiny (1.05$\pm$0.08)E+0	&\tiny 3.0E-2	&\tiny (1.85$\pm$0.60)E-1	\\
\tiny	020819B	&\tiny 14:57:38.600	&\tiny (9.20$\pm$0.75)E+0	&\tiny 4.0E-1	&\tiny -(2.13$\pm$0.64)E-1	\\
\tiny	020828	&\tiny 05:45:37.925	&\tiny (6.60$\pm$0.45)E-1	&\tiny 3.0E-2	&\tiny (3.41$\pm$0.48)E-1	\\
\tiny	020910	&\tiny 19:57:43.150	&\tiny (2.85$\pm$0.10)E+1	&\tiny 3.0E-1	&\tiny (1.33$\pm$0.29)E-1	\\
\tiny	020914	&\tiny 21:53:20.100	&\tiny (1.08$\pm$0.09)E+1	&\tiny 2.0E-1	&\tiny -(2.54$\pm$0.73)E-1	\\
\tiny	020926	&\tiny 04:52:54.200	&\tiny (2.24$\pm$0.15)E+1	&\tiny 8.0E-1	&\tiny -(2.05$\pm$0.53)E-1	\\
\tiny	021008A	&\tiny 07:01:03.550	&\tiny (1.45$\pm$0.01)E+1	&\tiny 1.0E-1	&\tiny (1.34$\pm$0.03)E-1	\\
\tiny	021008B	&\tiny 14:30:04.500	&\tiny (1.42$\pm$0.12)E+1	&\tiny 2.0E-1	&\tiny (0.09$\pm$6.92)E-2	\\
\tiny	021011	&\tiny 04:38:11.250	&\tiny (5.10$\pm$0.32)E+1	&\tiny 5.0E-1	&\tiny -(2.94$\pm$0.61)E-1	\\
\tiny	021016	&\tiny 10:29:43.500	&\tiny (8.30$\pm$0.54)E+1	&\tiny 1.0E+0	&\tiny -(1.54$\pm$0.55)E-1	\\
\tiny	021020	&\tiny 20:12:57.350	&\tiny (1.44$\pm$0.04)E+1	&\tiny 3.0E-1	&\tiny -(0.64$\pm$1.86)E-2	\\
\tiny	021023	&\tiny 02:53:47.300	&\tiny (1.36$\pm$0.05)E+1	&\tiny 2.0E-1	&\tiny -(0.60$\pm$2.85)E-2	\\
\tiny	021025	&\tiny 20:18:30.150	&\tiny (1.86$\pm$0.22)E+1	&\tiny 3.0E-1	&\tiny -(1.22$\pm$0.95)E-1	\\
\tiny	021102	&\tiny 15:58:31.850	&\tiny (1.08$\pm$0.05)E+1	&\tiny 3.0E-1	&\tiny -(4.04$\pm$0.41)E-1	\\
\tiny	021105	&\tiny 05:27:18.650	&\tiny (6.40$\pm$0.80)E+0	&\tiny 1.0E-1	&\tiny -(9.92$\pm$9.89)E-2	\\
\tiny	021108	&\tiny 05:39:55.800	&\tiny (2.20$\pm$0.17)E+1	&\tiny 4.0E-1	&\tiny -(5.99$\pm$1.07)E-1	\\
\tiny	021109	&\tiny 08:42:49.000	&\tiny (1.96$\pm$0.13)E+1	&\tiny 4.0E-1	&\tiny -(2.87$\pm$5.23)E-2	\\
\tiny	021113	&\tiny 13:37:37.625	&\tiny (3.98$\pm$0.48)E+1	&\tiny 2.5E-1	&\tiny -(2.82$\pm$1.08)E-1	\\
\tiny	021115	&\tiny 13:33:04.250	&\tiny (1.26$\pm$0.15)E+1	&\tiny 3.0E-1	&\tiny -(4.79$\pm$1.31)E-1	\\
\tiny	021119	&\tiny 12:54:07.700	&\tiny (2.82$\pm$0.08)E+1	&\tiny 2.0E-1	&\tiny -(1.83$\pm$0.26)E-1	\\
\tiny	021125	&\tiny 17:58:31.250	&\tiny (6.75$\pm$0.12)E+1	&\tiny 5.0E-1	&\tiny -(3.57$\pm$0.18)E-1	\\
\tiny	021201	&\tiny 05:30:04.175	&\tiny (3.40$\pm$0.16)E-1	&\tiny 1.0E-2	&\tiny (4.03$\pm$0.36)E-1	\\
\tiny	021205	&\tiny 03:18:29.750	&\tiny (7.65$\pm$0.33)E+1	&\tiny 1.5E+0	&\tiny -(2.23$\pm$0.36)E-1	\\
\tiny	021206	&\tiny 22:49:16.650	&\tiny (4.92$\pm$0.02)E+0	&\tiny 2.0E-2	&\tiny (2.26$\pm$0.22)E-2	\\
\tiny	021211	&\tiny 11:18:35.040	&\tiny (4.32$\pm$0.32)E+0	&\tiny 8.0E-2	&\tiny -(2.79$\pm$0.68)E-1	\\
\tiny	021214	&\tiny 03:27:25.500	&\tiny (3.00$\pm$0.28)E+1	&\tiny 1.0E+0	&\tiny -(2.07$\pm$0.77)E-1	\\
\tiny	021223	&\tiny 01:10:00.250	&\tiny (8.00$\pm$0.64)E+0	&\tiny 1.0E-1	&\tiny -(2.92$\pm$0.77)E-1	\\
\tiny	021226	&\tiny 14:53:39.675	&\tiny (3.50$\pm$0.53)E-1	&\tiny 5.0E-2	&\tiny (3.34$\pm$0.43)E-1	\\
\tiny	030102	&\tiny 23:18:59.350	&\tiny (1.32$\pm$0.09)E+1	&\tiny 3.0E-1	&\tiny -(1.65$\pm$0.56)E-1	\\

\end{tabular}
\end{table}
\end{center}

\begin{center}
\begin{table} \caption{The RHESSI GRB data-set.}
\centering
\begin{tabular}{lrrrr}

\hline
\\[-2ex]
GRB     & peak time         & $T_{90}$                          & $\delta t_{res}$ & hardness ratio         \\[0.3ex]
        & UTC               & [s]                               & [s]              & log $H$                \\[0.3ex]
\hline\hline
\\[-2ex]
\tiny	030103	&\tiny 21:46:46.750	&\tiny (1.00$\pm$0.12)E+1	&\tiny 5.0E-1	&\tiny -(3.93$\pm$1.14)E-1	\\
\tiny	030105	&\tiny 14:34:11.995	&\tiny (1.23$\pm$0.06)E+0	&\tiny 3.0E-2	&\tiny (3.77$\pm$0.40)E-1	\\
\tiny	030110	&\tiny 09:39:30.205	&\tiny (9.00$\pm$3.14)E-2	&\tiny 3.0E-2	&\tiny -(2.04$\pm$0.88)E-1	\\
\tiny	030115A	&\tiny 06:25:37.250	&\tiny (7.95$\pm$0.14)E+1	&\tiny 5.0E-1	&\tiny (3.54$\pm$1.40)E-2	\\
\tiny	030115B	&\tiny 08:15:48.500	&\tiny (2.70$\pm$0.10)E+1	&\tiny 2.0E-1	&\tiny -(2.83$\pm$0.36)E-1	\\
\tiny	030127	&\tiny 12:32:42.750	&\tiny (3.80$\pm$0.40)E+1	&\tiny 5.0E-1	&\tiny (2.34$\pm$8.30)E-2	\\
\tiny	030204	&\tiny 12:45:36.500	&\tiny (5.60$\pm$0.13)E+1	&\tiny 1.0E+0	&\tiny (1.39$\pm$1.34)E-2	\\
\tiny	030206	&\tiny 11:00:32.010	&\tiny (1.40$\pm$0.24)E-1	&\tiny 2.0E-2	&\tiny (2.62$\pm$0.83)E-1	\\
\tiny	030212	&\tiny 22:17:18.450	&\tiny (2.31$\pm$0.26)E+1	&\tiny 3.0E-1	&\tiny -(2.25$\pm$0.97)E-1	\\
\tiny	030214	&\tiny 14:48:21.325	&\tiny (1.83$\pm$0.10)E+1	&\tiny 1.5E-1	&\tiny -(2.07$\pm$0.49)E-1	\\
\tiny	030216	&\tiny 16:13:44.150	&\tiny (9.30$\pm$1.40)E+0	&\tiny 3.0E-1	&\tiny -(1.20$\pm$1.16)E-1	\\
\tiny	030217	&\tiny 02:45:42.600	&\tiny (5.40$\pm$0.08)E+1	&\tiny 8.0E-2	&\tiny (4.46$\pm$1.28)E-2	\\
\tiny	030222	&\tiny 16:32:53.250	&\tiny (1.44$\pm$0.18)E+1	&\tiny 3.0E-1	&\tiny -(2.01$\pm$1.02)E-1	\\
\tiny	030223	&\tiny 09:45:15.250	&\tiny (2.05$\pm$0.11)E+1	&\tiny 5.0E-1	&\tiny (2.50$\pm$0.42)E-1	\\
\tiny	030225	&\tiny 15:02:53.750	&\tiny (2.00$\pm$0.11)E+1	&\tiny 5.0E-1	&\tiny -(9.23$\pm$4.20)E-2	\\
\tiny	030227	&\tiny 13:12:05.250	&\tiny (7.10$\pm$1.03)E+1	&\tiny 5.0E-1	&\tiny (0.68$\pm$1.11)E-1	\\
\tiny	030228	&\tiny 20:26:47.250	&\tiny (2.80$\pm$0.18)E+1	&\tiny 5.0E-1	&\tiny -(3.96$\pm$4.98)E-2	\\
\tiny	030301	&\tiny 20:27:21.500	&\tiny (4.30$\pm$0.35)E+1	&\tiny 1.0E+0	&\tiny -(3.73$\pm$0.85)E-1	\\
\tiny	030306	&\tiny 03:38:16.150	&\tiny (1.26$\pm$0.04)E+1	&\tiny 3.0E-1	&\tiny -(1.16$\pm$0.22)E-1	\\
\tiny	030307	&\tiny 14:31:58.950	&\tiny (3.80$\pm$0.11)E+0	&\tiny 1.0E-1	&\tiny -(1.34$\pm$0.13)E-1	\\
\tiny	030320A	&\tiny 10:12:01.150	&\tiny (4.86$\pm$0.15)E+1	&\tiny 3.0E-1	&\tiny -(6.10$\pm$2.51)E-2	\\
\tiny	030320B	&\tiny 18:49:35.500	&\tiny (1.56$\pm$0.06)E+2	&\tiny 3.0E+0	&\tiny -(6.38$\pm$2.83)E-2	\\
\tiny	030326	&\tiny 10:43:44.150	&\tiny (1.26$\pm$0.06)E+1	&\tiny 3.0E-1	&\tiny -(2.01$\pm$0.34)E-1	\\
\tiny	030328	&\tiny 07:28:48.250	&\tiny (6.55$\pm$0.26)E+1	&\tiny 5.0E-1	&\tiny -(4.09$\pm$3.25)E-2	\\
\tiny	030329A	&\tiny 11:37:40.850	&\tiny (1.74$\pm$0.03)E+1	&\tiny 3.0E-1	&\tiny -(2.60$\pm$0.06)E-1	\\
\tiny	030329B	&\tiny 15:34:18.500	&\tiny (6.10$\pm$0.29)E+1	&\tiny 1.0E+0	&\tiny -(1.50$\pm$0.39)E-1	\\
\tiny	030331	&\tiny 05:38:49.500	&\tiny (2.38$\pm$0.11)E+1	&\tiny 2.0E-1	&\tiny -(3.34$\pm$0.47)E-1	\\
\tiny	030406	&\tiny 22:42:57.450	&\tiny (7.02$\pm$0.13)E+1	&\tiny 3.0E-1	&\tiny -(3.30$\pm$1.54)E-2	\\
\tiny	030410	&\tiny 11:23:42.275	&\tiny (1.60$\pm$0.16)E+0	&\tiny 5.0E-2	&\tiny (4.10$\pm$1.01)E-1	\\
\tiny	030413	&\tiny 07:34:45.550	&\tiny (2.04$\pm$0.09)E+1	&\tiny 3.0E-1	&\tiny (1.48$\pm$0.35)E-1	\\
\tiny	030414	&\tiny 13:48:28.250	&\tiny (2.85$\pm$0.08)E+1	&\tiny 5.0E-1	&\tiny -(3.66$\pm$1.80)E-2	\\
\tiny	030419	&\tiny 01:12:14.300	&\tiny (3.78$\pm$0.06)E+1	&\tiny 2.0E-1	&\tiny -(2.81$\pm$0.14)E-1	\\
\tiny	030421	&\tiny 00:36:32.350	&\tiny (1.27$\pm$0.04)E+1	&\tiny 1.0E-1	&\tiny (8.23$\pm$2.40)E-2	\\
\tiny	030422	&\tiny 09:01:30.850	&\tiny (1.47$\pm$0.11)E+1	&\tiny 3.0E-1	&\tiny -(3.83$\pm$6.08)E-2	\\
\tiny	030428	&\tiny 22:31:23.300	&\tiny (9.60$\pm$0.26)E+0	&\tiny 2.0E-1	&\tiny (5.70$\pm$1.42)E-2	\\
\tiny	030501A	&\tiny 01:17:22.300	&\tiny (7.40$\pm$0.49)E+0	&\tiny 2.0E-1	&\tiny -(2.21$\pm$0.55)E-1	\\
\tiny	030501B	&\tiny 03:10:23.850	&\tiny (1.17$\pm$0.16)E+1	&\tiny 3.0E-1	&\tiny -(4.00$\pm$1.43)E-1	\\
\tiny	030501C	&\tiny 20:44:49.520	&\tiny (1.24$\pm$0.12)E+0	&\tiny 4.0E-2	&\tiny (2.21$\pm$0.74)E-1	\\
\tiny	030505A	&\tiny 07:39:13.100	&\tiny (8.82$\pm$0.43)E+1	&\tiny 6.0E-1	&\tiny (3.77$\pm$3.96)E-2	\\
\tiny	030505B	&\tiny 09:03:26.750	&\tiny (1.07$\pm$0.06)E+2	&\tiny 1.5E+0	&\tiny -(6.53$\pm$4.77)E-2	\\
\tiny	030506	&\tiny 02:04:27.350	&\tiny (2.31$\pm$0.07)E+1	&\tiny 3.0E-1	&\tiny -(9.31$\pm$2.34)E-2	\\
\tiny	030518A	&\tiny 01:23:49.650	&\tiny (2.52$\pm$0.02)E+1	&\tiny 1.0E-1	&\tiny (4.79$\pm$0.07)E-1	\\
\tiny	030518B	&\tiny 03:12:23.050	&\tiny (1.74$\pm$0.12)E+1	&\tiny 3.0E-1	&\tiny (3.98$\pm$0.54)E-1	\\
\tiny	030519A	&\tiny 09:32:22.700	&\tiny (3.20$\pm$0.39)E+0	&\tiny 2.0E-1	&\tiny (6.78$\pm$1.13)E-1	\\
\tiny	030519B	&\tiny 14:05:00.100	&\tiny (9.00$\pm$0.21)E+0	&\tiny 2.0E-1	&\tiny -(1.81$\pm$0.47)E-2	\\
\tiny	030523	&\tiny 14:10:53.190	&\tiny (1.20$\pm$0.26)E-1	&\tiny 2.0E-2	&\tiny (0.22$\pm$1.04)E-1	\\
\tiny	030528	&\tiny 13:03:07.750	&\tiny (2.15$\pm$0.17)E+1	&\tiny 5.0E-1	&\tiny -(2.26$\pm$0.69)E-1	\\
\tiny	030601	&\tiny 22:12:06.050	&\tiny (1.83$\pm$0.08)E+1	&\tiny 3.0E-1	&\tiny (7.59$\pm$3.59)E-2	\\
\tiny	030614	&\tiny 01:30:42.500	&\tiny (1.68$\pm$0.05)E+2	&\tiny 3.0E+0	&\tiny -(2.19$\pm$0.21)E-1	\\
\tiny	030626	&\tiny 01:46:54.950	&\tiny (3.85$\pm$0.10)E+1	&\tiny 7.0E-1	&\tiny (2.46$\pm$0.18)E-1	\\
\tiny	030703	&\tiny 19:14:02.400	&\tiny (5.68$\pm$0.74)E+1	&\tiny 8.0E-1	&\tiny -(0.84$\pm$1.03)E-1	\\
\tiny	030706	&\tiny 00:02:17.750	&\tiny (4.50$\pm$0.45)E+0	&\tiny 3.0E-1	&\tiny (1.36$\pm$0.62)E-1	\\
\tiny	030710	&\tiny 23:05:02.350	&\tiny (1.77$\pm$0.07)E+1	&\tiny 3.0E-1	&\tiny -(7.49$\pm$0.71)E-1	\\
\tiny	030714	&\tiny 22:14:50.300	&\tiny (1.02$\pm$0.08)E+1	&\tiny 2.0E-1	&\tiny -(2.73$\pm$0.73)E-1	\\
\tiny	030716	&\tiny 11:57:19.750	&\tiny (5.35$\pm$0.40)E+1	&\tiny 5.0E-1	&\tiny -(3.16$\pm$0.75)E-1	\\
\tiny	030721	&\tiny 23:41:10.100	&\tiny (2.96$\pm$0.05)E+1	&\tiny 2.0E-1	&\tiny (2.88$\pm$0.14)E-1	\\
\tiny	030725	&\tiny 11:46:27.800	&\tiny (1.84$\pm$0.07)E+1	&\tiny 4.0E-1	&\tiny -(2.22$\pm$0.32)E-1	\\
\tiny	030726A	&\tiny 06:38:41.750	&\tiny (2.75$\pm$0.06)E+1	&\tiny 5.0E-1	&\tiny (1.56$\pm$0.11)E-1	\\
\tiny	030726B	&\tiny 09:51:37.500	&\tiny (1.94$\pm$0.12)E+2	&\tiny 1.0E+0	&\tiny -(2.91$\pm$0.63)E-1	\\
\tiny	030728	&\tiny 09:05:55.100	&\tiny (1.92$\pm$0.28)E+1	&\tiny 2.0E-1	&\tiny (4.26$\pm$1.50)E-1	\\
\tiny	030824	&\tiny 06:31:10.850	&\tiny (3.00$\pm$0.20)E+1	&\tiny 3.0E-1	&\tiny -(6.11$\pm$0.95)E-1	\\
\tiny	030827	&\tiny 16:08:40.325	&\tiny (1.85$\pm$0.08)E+0	&\tiny 5.0E-2	&\tiny (7.30$\pm$2.77)E-2	\\
\tiny	030830	&\tiny 18:37:46.250	&\tiny (2.05$\pm$0.09)E+1	&\tiny 5.0E-1	&\tiny -(1.27$\pm$0.34)E-1	\\
\tiny	030831	&\tiny 15:07:22.450	&\tiny (2.31$\pm$0.09)E+1	&\tiny 3.0E-1	&\tiny -(1.04$\pm$0.33)E-1	\\
\tiny	030919	&\tiny 21:10:36.750	&\tiny (2.00$\pm$0.14)E+1	&\tiny 5.0E-1	&\tiny -(2.05$\pm$0.59)E-1	\\
\tiny	030921	&\tiny 08:38:23.650	&\tiny (1.47$\pm$0.06)E+1	&\tiny 3.0E-1	&\tiny (2.23$\pm$3.15)E-2	\\

\end{tabular}
\end{table}
\end{center}

\begin{center}
\begin{table} \caption{The RHESSI GRB data-set.}
\centering
\begin{tabular}{lrrrr}

\hline
\\[-2ex]
GRB     & peak time         & $T_{90}$                          & $\delta t_{res}$ & hardness ratio         \\[0.3ex]
        & UTC               & [s]                               & [s]              & log $H$                \\[0.3ex]
\hline\hline
\\[-2ex]
\tiny	030922A	&\tiny 08:43:35.950	&\tiny (2.28$\pm$0.08)E+1	&\tiny 3.0E-1	&\tiny -(1.07$\pm$0.30)E-1	\\
\tiny	030922B	&\tiny 18:30:56.900	&\tiny (1.08$\pm$0.03)E+1	&\tiny 2.0E-1	&\tiny (4.63$\pm$1.41)E-2	\\
\tiny	030926	&\tiny 16:52:28.290	&\tiny (2.80$\pm$0.31)E-1	&\tiny 2.0E-2	&\tiny (2.89$\pm$0.77)E-1	\\
\tiny	031005	&\tiny 08:21:50.250	&\tiny (1.60$\pm$0.18)E+1	&\tiny 5.0E-1	&\tiny -(4.44$\pm$1.19)E-1	\\
\tiny	031019	&\tiny 22:00:57.175	&\tiny (7.50$\pm$0.71)E+0	&\tiny 1.5E-1	&\tiny (2.48$\pm$0.79)E-1	\\
\tiny	031024	&\tiny 09:24:14.350	&\tiny (4.00$\pm$0.32)E+0	&\tiny 1.0E-1	&\tiny (1.77$\pm$0.58)E-1	\\
\tiny	031027	&\tiny 17:07:50.250	&\tiny (3.90$\pm$0.07)E+1	&\tiny 5.0E-1	&\tiny -(1.03$\pm$0.10)E-1	\\
\tiny	031107	&\tiny 18:24:07.400	&\tiny (1.92$\pm$0.06)E+1	&\tiny 4.0E-1	&\tiny (1.02$\pm$2.13)E-2	\\
\tiny	031108	&\tiny 14:11:19.250	&\tiny (2.75$\pm$0.07)E+1	&\tiny 5.0E-1	&\tiny (1.67$\pm$0.16)E-1	\\
\tiny	031111	&\tiny 16:45:20.920	&\tiny (3.20$\pm$0.08)E+0	&\tiny 8.0E-2	&\tiny (1.50$\pm$0.08)E-1	\\
\tiny	031118	&\tiny 06:26:16.410	&\tiny (2.60$\pm$0.27)E-1	&\tiny 2.0E-2	&\tiny (2.01$\pm$0.57)E-1	\\
\tiny	031120	&\tiny 05:52:38.250	&\tiny (9.75$\pm$0.30)E+1	&\tiny 1.5E+0	&\tiny (4.24$\pm$2.27)E-2	\\
\tiny	031127	&\tiny 18:58:55.050	&\tiny (1.92$\pm$0.17)E+1	&\tiny 3.0E-1	&\tiny -(2.88$\pm$0.83)E-1	\\
\tiny	031130	&\tiny 02:04:53.775	&\tiny (4.65$\pm$0.29)E+0	&\tiny 1.5E-1	&\tiny (5.52$\pm$4.44)E-2	\\
\tiny	031214	&\tiny 22:50:44.150	&\tiny (8.30$\pm$1.01)E+0	&\tiny 1.0E-1	&\tiny -(6.98$\pm$9.55)E-2	\\
\tiny	031218	&\tiny 06:28:09.370	&\tiny (2.20$\pm$0.24)E-1	&\tiny 2.0E-2	&\tiny (1.46$\pm$0.51)E-1	\\
\tiny	031219	&\tiny 05:39:05.850	&\tiny (7.50$\pm$0.21)E+0	&\tiny 1.0E-1	&\tiny -(1.21$\pm$0.21)E-1	\\
\tiny	031226A	&\tiny 15:43:08.750	&\tiny (3.45$\pm$0.29)E+1	&\tiny 5.0E-1	&\tiny -(4.63$\pm$1.03)E-1	\\
\tiny	031226B	&\tiny 17:51:29.750	&\tiny (4.40$\pm$0.23)E+1	&\tiny 5.0E-1	&\tiny -(5.74$\pm$4.22)E-2	\\
\tiny	040102	&\tiny 19:35:26.975	&\tiny (7.80$\pm$0.83)E+0	&\tiny 1.5E-1	&\tiny -(4.43$\pm$1.15)E-1	\\
\tiny	040108	&\tiny 07:46:39.900	&\tiny (3.30$\pm$0.37)E+1	&\tiny 6.0E-1	&\tiny -(2.64$\pm$1.03)E-1	\\
\tiny	040113	&\tiny 01:36:46.350	&\tiny (1.62$\pm$0.10)E+1	&\tiny 3.0E-1	&\tiny -(4.60$\pm$5.01)E-2	\\
\tiny	040115	&\tiny 18:30:11.550	&\tiny (2.19$\pm$0.24)E+1	&\tiny 3.0E-1	&\tiny -(2.40$\pm$0.94)E-1	\\
\tiny	040125	&\tiny 22:14:47.100	&\tiny (1.60$\pm$0.23)E+1	&\tiny 2.0E-1	&\tiny -(1.67$\pm$1.16)E-1	\\
\tiny	040205A	&\tiny 05:19:47.350	&\tiny (3.90$\pm$0.62)E+0	&\tiny 1.0E-1	&\tiny -(2.06$\pm$1.29)E-1	\\
\tiny	040205B	&\tiny 09:27:45.750	&\tiny (2.65$\pm$0.33)E+1	&\tiny 5.0E-1	&\tiny (2.19$\pm$0.98)E-1	\\
\tiny	040207	&\tiny 22:12:22.600	&\tiny (2.44$\pm$0.06)E+1	&\tiny 4.0E-1	&\tiny (1.07$\pm$0.17)E-1	\\
\tiny	040211	&\tiny 15:02:08.050	&\tiny (3.70$\pm$0.52)E+0	&\tiny 1.0E-1	&\tiny -(2.61$\pm$1.25)E-1	\\
\tiny	040215	&\tiny 00:28:02.500	&\tiny (5.00$\pm$0.29)E+1	&\tiny 1.0E+0	&\tiny -(3.57$\pm$0.57)E-1	\\
\tiny	040220	&\tiny 00:55:15.800	&\tiny (1.72$\pm$0.09)E+1	&\tiny 4.0E-1	&\tiny (5.71$\pm$0.46)E-1	\\
\tiny	040225A	&\tiny 05:30:53.850	&\tiny (1.05$\pm$0.11)E+1	&\tiny 3.0E-1	&\tiny -(8.84$\pm$8.12)E-2	\\
\tiny	040225B	&\tiny 10:02:12.200	&\tiny (1.40$\pm$0.16)E+1	&\tiny 4.0E-1	&\tiny -(3.12$\pm$1.10)E-1	\\
\tiny	040228	&\tiny 00:09:08.500	&\tiny (2.73$\pm$0.02)E+2	&\tiny 2.0E-1	&\tiny -(1.58$\pm$0.07)E-1	\\
\tiny	040302A	&\tiny 04:14:35.250	&\tiny (2.37$\pm$0.27)E+1	&\tiny 3.0E-1	&\tiny -(2.76$\pm$1.05)E-1	\\
\tiny	040302B	&\tiny 12:24:03.900	&\tiny (1.08$\pm$0.02)E+1	&\tiny 2.0E-1	&\tiny -(7.71$\pm$0.64)E-2	\\
\tiny	040303	&\tiny 15:32:37.750	&\tiny (2.97$\pm$0.27)E+1	&\tiny 3.0E-1	&\tiny -(1.52$\pm$0.76)E-1	\\
\tiny	040312	&\tiny 00:02:36.550	&\tiny (1.60$\pm$0.22)E-1	&\tiny 2.0E-2	&\tiny (2.13$\pm$0.48)E-1	\\
\tiny	040316	&\tiny 18:16:14.350	&\tiny (1.50$\pm$0.02)E+1	&\tiny 1.0E-1	&\tiny -(4.87$\pm$1.30)E-2	\\
\tiny	040323	&\tiny 13:03:04.750	&\tiny (1.89$\pm$0.19)E+1	&\tiny 3.0E-1	&\tiny -(2.92$\pm$0.94)E-1	\\
\tiny	040324	&\tiny 10:21:12.908	&\tiny (2.55$\pm$0.17)E-1	&\tiny 1.5E-2	&\tiny (2.54$\pm$0.26)E-1	\\
\tiny	040327	&\tiny 16:19:27.500	&\tiny (2.00$\pm$0.32)E+1	&\tiny 1.0E+0	&\tiny -(4.06$\pm$1.54)E-1	\\
\tiny	040329	&\tiny 11:10:51.965	&\tiny (2.07$\pm$0.04)E+0	&\tiny 3.0E-2	&\tiny (3.06$\pm$0.12)E-1	\\
\tiny	040330	&\tiny 13:14:42.150	&\tiny (2.88$\pm$0.41)E+1	&\tiny 3.0E-1	&\tiny (0.04$\pm$1.09)E-1	\\
\tiny	040404	&\tiny 10:58:52.150	&\tiny (4.90$\pm$0.35)E+0	&\tiny 1.0E-1	&\tiny -(1.38$\pm$0.60)E-1	\\
\tiny	040413	&\tiny 13:09:56.490	&\tiny (2.80$\pm$0.31)E-1	&\tiny 2.0E-2	&\tiny (1.21$\pm$0.68)E-1	\\
\tiny	040414	&\tiny 11:09:22.500	&\tiny (7.70$\pm$0.24)E+1	&\tiny 1.0E+0	&\tiny -(1.63$\pm$2.36)E-2	\\
\tiny	040421	&\tiny 02:30:27.300	&\tiny (1.14$\pm$0.02)E+1	&\tiny 2.0E-1	&\tiny (2.73$\pm$0.08)E-1	\\
\tiny	040423	&\tiny 02:23:30.350	&\tiny (4.20$\pm$0.53)E+0	&\tiny 3.0E-1	&\tiny -(1.16$\pm$0.87)E-1	\\
\tiny	040425	&\tiny 16:23:34.275	&\tiny (8.10$\pm$0.19)E+0	&\tiny 5.0E-2	&\tiny (1.08$\pm$0.20)E-1	\\
\tiny	040427	&\tiny 20:12:37.700	&\tiny (8.00$\pm$0.77)E+0	&\tiny 2.0E-1	&\tiny -(4.55$\pm$7.43)E-2	\\
\tiny	040429	&\tiny 10:53:04.400	&\tiny (2.52$\pm$0.19)E+1	&\tiny 4.0E-1	&\tiny (2.85$\pm$6.00)E-2	\\
\tiny	040502A	&\tiny 06:37:06.900	&\tiny (1.86$\pm$0.04)E+1	&\tiny 2.0E-1	&\tiny -(2.97$\pm$0.22)E-1	\\
\tiny	040502B	&\tiny 13:30:02.500	&\tiny (1.66$\pm$0.07)E+2	&\tiny 1.0E+0	&\tiny -(2.32$\pm$0.39)E-1	\\
\tiny	040506	&\tiny 23:45:18.800	&\tiny (6.56$\pm$0.42)E+1	&\tiny 8.0E-1	&\tiny -(2.34$\pm$0.58)E-1	\\
\tiny	040508	&\tiny 10:15:44.750	&\tiny (4.05$\pm$0.56)E+1	&\tiny 5.0E-1	&\tiny -(0.95$\pm$1.08)E-1	\\
\tiny	040510	&\tiny 09:59:37.300	&\tiny (6.00$\pm$0.57)E+0	&\tiny 2.0E-1	&\tiny -(2.87$\pm$0.83)E-1	\\
\tiny	040513	&\tiny 03:02:17.500	&\tiny (5.30$\pm$0.83)E+1	&\tiny 1.0E+0	&\tiny -(0.78$\pm$1.20)E-1	\\
\tiny	040526	&\tiny 20:21:13.350	&\tiny (1.29$\pm$0.19)E+1	&\tiny 3.0E-1	&\tiny -(4.19$\pm$1.53)E-1	\\
\tiny	040528	&\tiny 16:55:58.150	&\tiny (2.16$\pm$0.07)E+1	&\tiny 3.0E-1	&\tiny -(1.06$\pm$0.26)E-1	\\
\tiny	040531	&\tiny 23:15:04.750	&\tiny (4.20$\pm$0.33)E+1	&\tiny 5.0E-1	&\tiny -(4.73$\pm$0.91)E-1	\\
\tiny	040601	&\tiny 06:33:24.625	&\tiny (2.73$\pm$0.29)E+1	&\tiny 2.5E-1	&\tiny (5.10$\pm$8.45)E-2	\\
\tiny	040603A	&\tiny 15:40:58.700	&\tiny (1.24$\pm$0.13)E+1	&\tiny 2.0E-1	&\tiny -(1.42$\pm$0.84)E-1	\\
\tiny	040603B	&\tiny 19:15:38.500	&\tiny (7.50$\pm$1.00)E+1	&\tiny 3.0E+0	&\tiny -(5.04$\pm$1.49)E-1	\\
\tiny	040605A	&\tiny 04:31:44.500	&\tiny (5.80$\pm$0.73)E+0	&\tiny 2.0E-1	&\tiny (3.25$\pm$1.04)E-1	\\
\tiny	040605B	&\tiny 18:46:16.390	&\tiny (1.80$\pm$0.22)E-1	&\tiny 2.0E-2	&\tiny (2.33$\pm$0.48)E-1	\\
\tiny	040605C	&\tiny 23:58:45.800	&\tiny (1.68$\pm$0.08)E+1	&\tiny 4.0E-1	&\tiny -(4.30$\pm$0.50)E-1	\\

\end{tabular}
\end{table}
\end{center}

\begin{center}
\begin{table} \caption{The RHESSI GRB data-set.}
\centering
\begin{tabular}{lrrrr}

\hline
\\[-2ex]
GRB     & peak time         & $T_{90}$                          & $\delta t_{res}$ & hardness ratio         \\[0.3ex]
        & UTC               & [s]                               & [s]              & log $H$                \\[0.3ex]
\hline\hline
\\[-2ex]
\tiny	040611	&\tiny 13:36:00.600	&\tiny (1.80$\pm$0.10)E+1	&\tiny 4.0E-1	&\tiny (2.02$\pm$0.44)E-1	\\
\tiny	040619	&\tiny 15:15:52.250	&\tiny (7.70$\pm$0.27)E+0	&\tiny 1.0E-1	&\tiny -(1.74$\pm$0.28)E-1	\\
\tiny	040701	&\tiny 22:46:45.250	&\tiny (9.60$\pm$0.69)E+0	&\tiny 3.0E-1	&\tiny (9.16$\pm$5.34)E-2	\\
\tiny	040719	&\tiny 01:16:31.250	&\tiny (7.50$\pm$0.65)E+0	&\tiny 1.0E-1	&\tiny -(1.77$\pm$0.73)E-1	\\
\tiny	040723	&\tiny 04:06:38.150	&\tiny (1.02$\pm$0.04)E+1	&\tiny 1.0E-1	&\tiny -(1.51$\pm$0.27)E-1	\\
\tiny	040731	&\tiny 10:24:42.750	&\tiny (2.45$\pm$0.07)E+1	&\tiny 5.0E-1	&\tiny (1.23$\pm$0.18)E-1	\\
\tiny	040803	&\tiny 15:08:57.750	&\tiny (1.14$\pm$0.11)E+2	&\tiny 1.5E+0	&\tiny -(3.95$\pm$1.05)E-1	\\
\tiny	040810	&\tiny 14:15:42.650	&\tiny (1.92$\pm$0.05)E+1	&\tiny 3.0E-1	&\tiny -(7.96$\pm$1.96)E-2	\\
\tiny	040818	&\tiny 01:29:04.250	&\tiny (8.00$\pm$0.33)E+0	&\tiny 1.0E-1	&\tiny (4.57$\pm$0.42)E-1	\\
\tiny	040822	&\tiny 21:21:55.125	&\tiny (1.38$\pm$0.14)E+0	&\tiny 3.0E-2	&\tiny (2.05$\pm$0.83)E-1	\\
\tiny	040824	&\tiny 05:16:07.500	&\tiny (4.90$\pm$0.50)E+1	&\tiny 1.0E+0	&\tiny -(4.90$\pm$7.92)E-2	\\
\tiny	040921	&\tiny 16:06:20.265	&\tiny (2.80$\pm$0.73)E-1	&\tiny 7.0E-2	&\tiny (1.88$\pm$0.64)E-1	\\
\tiny	040925	&\tiny 22:28:59.750	&\tiny (4.75$\pm$0.16)E+1	&\tiny 5.0E-1	&\tiny (6.66$\pm$2.68)E-2	\\
\tiny	040926	&\tiny 04:03:18.100	&\tiny (7.60$\pm$0.22)E+0	&\tiny 2.0E-1	&\tiny (1.93$\pm$0.11)E-1	\\
\tiny	041003	&\tiny 09:17:56.250	&\tiny (9.00$\pm$0.74)E+0	&\tiny 5.0E-1	&\tiny -(1.16$\pm$0.52)E-1	\\
\tiny	041006	&\tiny 12:18:40.350	&\tiny (1.11$\pm$0.08)E+1	&\tiny 3.0E-1	&\tiny -(4.71$\pm$0.80)E-1	\\
\tiny	041007	&\tiny 02:02:08.350	&\tiny (2.00$\pm$0.13)E+0	&\tiny 1.0E-1	&\tiny (3.52$\pm$0.37)E-1	\\
\tiny	041009	&\tiny 06:38:21.250	&\tiny (7.50$\pm$0.76)E+0	&\tiny 3.0E-1	&\tiny -(8.60$\pm$7.65)E-2	\\
\tiny	041010	&\tiny 00:14:57.555	&\tiny (2.50$\pm$0.22)E-1	&\tiny 1.0E-2	&\tiny (1.58$\pm$0.64)E-1	\\
\tiny	041012	&\tiny 12:40:51.500	&\tiny (4.60$\pm$0.53)E+1	&\tiny 1.0E+0	&\tiny -(3.90$\pm$1.16)E-1	\\
\tiny	041013A	&\tiny 02:35:25.000	&\tiny (1.84$\pm$0.09)E+2	&\tiny 2.0E+0	&\tiny (1.27$\pm$3.84)E-2	\\
\tiny	041013B	&\tiny 22:56:27.865	&\tiny (3.60$\pm$0.46)E-1	&\tiny 3.0E-2	&\tiny (6.92$\pm$7.69)E-2	\\
\tiny	041015	&\tiny 10:22:18.950	&\tiny (3.90$\pm$0.29)E+0	&\tiny 1.0E-1	&\tiny -(1.82$\pm$0.60)E-1	\\
\tiny	041016	&\tiny 04:39:37.000	&\tiny (1.80$\pm$0.17)E+1	&\tiny 4.0E-1	&\tiny -(1.90$\pm$0.82)E-1	\\
\tiny	041018	&\tiny 13:08:19.500	&\tiny (1.02$\pm$0.04)E+2	&\tiny 1.0E+0	&\tiny -(3.19$\pm$3.19)E-2	\\
\tiny	041101	&\tiny 01:49:36.150	&\tiny (3.20$\pm$0.34)E+0	&\tiny 1.0E-1	&\tiny (2.28$\pm$0.85)E-1	\\
\tiny	041102	&\tiny 11:12:23.750	&\tiny (2.70$\pm$0.22)E+0	&\tiny 1.0E-1	&\tiny (1.69$\pm$0.60)E-1	\\
\tiny	041107	&\tiny 15:49:29.250	&\tiny (4.70$\pm$0.45)E+1	&\tiny 5.0E-1	&\tiny -(6.04$\pm$7.68)E-2	\\
\tiny	041116	&\tiny 05:34:56.500	&\tiny (5.00$\pm$0.53)E+1	&\tiny 1.0E+0	&\tiny -(3.38$\pm$1.06)E-1	\\
\tiny	041117	&\tiny 15:18:00.950	&\tiny (1.62$\pm$0.05)E+1	&\tiny 3.0E-1	&\tiny -(1.81$\pm$0.24)E-1	\\
\tiny	041120	&\tiny 19:23:41.300	&\tiny (8.60$\pm$0.42)E+0	&\tiny 2.0E-1	&\tiny -(3.20$\pm$0.44)E-1	\\
\tiny	041125	&\tiny 16:07:27.400	&\tiny (2.52$\pm$0.04)E+1	&\tiny 4.0E-1	&\tiny -(2.34$\pm$0.06)E-1	\\
\tiny	041202	&\tiny 02:30:57.300	&\tiny (1.66$\pm$0.03)E+1	&\tiny 2.0E-1	&\tiny (1.14$\pm$0.12)E-1	\\
\tiny	041211A	&\tiny 07:49:56.450	&\tiny (2.06$\pm$0.05)E+1	&\tiny 1.0E-1	&\tiny -(1.33$\pm$0.23)E-1	\\
\tiny	041211B	&\tiny 11:31:53.400	&\tiny (2.00$\pm$0.09)E+1	&\tiny 4.0E-1	&\tiny (8.20$\pm$3.24)E-2	\\
\tiny	041211C	&\tiny 23:57:42.925	&\tiny (6.15$\pm$0.16)E+0	&\tiny 1.5E-1	&\tiny (4.46$\pm$0.07)E-1	\\
\tiny	041213	&\tiny 06:59:36.330	&\tiny (1.40$\pm$0.24)E-1	&\tiny 2.0E-2	&\tiny (3.19$\pm$0.82)E-1	\\
\tiny	041218	&\tiny 15:45:50.500	&\tiny (5.10$\pm$0.57)E+1	&\tiny 1.0E+0	&\tiny -(2.22$\pm$0.98)E-1	\\
\tiny	041219	&\tiny 01:42:19.400	&\tiny (1.00$\pm$0.11)E+1	&\tiny 4.0E-1	&\tiny -(1.32$\pm$7.84)E-2	\\
\tiny	041223	&\tiny 14:06:42.250	&\tiny (4.05$\pm$0.11)E+1	&\tiny 5.0E-1	&\tiny (8.55$\pm$2.11)E-2	\\
\tiny	041224	&\tiny 20:20:58.250	&\tiny (3.85$\pm$0.58)E+1	&\tiny 5.0E-1	&\tiny -(3.21$\pm$1.42)E-1	\\
\tiny	041231	&\tiny 21:50:48.050	&\tiny (1.00$\pm$0.12)E+0	&\tiny 1.0E-1	&\tiny (3.20$\pm$0.56)E-1	\\
\tiny	050124	&\tiny 11:30:03.250	&\tiny (2.80$\pm$0.38)E+0	&\tiny 1.0E-1	&\tiny -(2.24$\pm$1.12)E-1	\\
\tiny	050126	&\tiny 21:07:37.700	&\tiny (3.30$\pm$0.10)E+1	&\tiny 6.0E-1	&\tiny -(1.71$\pm$0.22)E-1	\\
\tiny	050203	&\tiny 17:22:00.850	&\tiny (3.80$\pm$0.24)E+0	&\tiny 1.0E-1	&\tiny -(0.85$\pm$4.62)E-2	\\
\tiny	050213	&\tiny 19:24:04.750	&\tiny (1.70$\pm$0.09)E+1	&\tiny 5.0E-1	&\tiny (2.43$\pm$0.37)E-1	\\
\tiny	050214	&\tiny 11:38:33.500	&\tiny (4.00$\pm$0.66)E+1	&\tiny 1.0E+0	&\tiny -(0.16$\pm$1.23)E-1	\\
\tiny	050216	&\tiny 07:26:34.275	&\tiny (5.00$\pm$0.72)E-1	&\tiny 5.0E-2	&\tiny (5.66$\pm$1.00)E-1	\\
\tiny	050219	&\tiny 21:05:51.650	&\tiny (9.40$\pm$0.17)E+0	&\tiny 1.0E-1	&\tiny -(3.01$\pm$0.15)E-1	\\
\tiny	050311	&\tiny 17:06:58.850	&\tiny (2.40$\pm$0.44)E+0	&\tiny 3.0E-1	&\tiny -(1.44$\pm$1.09)E-1	\\
\tiny	050312	&\tiny 05:40:13.575	&\tiny (1.50$\pm$0.51)E-1	&\tiny 5.0E-2	&\tiny (3.35$\pm$0.82)E-1	\\
\tiny	050314	&\tiny 08:33:08.900	&\tiny (8.00$\pm$0.35)E+0	&\tiny 2.0E-1	&\tiny -(4.77$\pm$0.45)E-1	\\
\tiny	050320	&\tiny 08:04:26.900	&\tiny (1.54$\pm$0.13)E+1	&\tiny 2.0E-1	&\tiny (5.51$\pm$6.52)E-2	\\
\tiny	050321	&\tiny 22:11:51.500	&\tiny (7.00$\pm$0.75)E+0	&\tiny 2.0E-1	&\tiny -(8.21$\pm$8.41)E-2	\\
\tiny	050326	&\tiny 09:53:56.500	&\tiny (2.70$\pm$0.11)E+1	&\tiny 2.0E-1	&\tiny -(2.54$\pm$0.36)E-1	\\
\tiny	050328	&\tiny 03:25:14.875	&\tiny (4.50$\pm$0.51)E-1	&\tiny 5.0E-2	&\tiny -(1.02$\pm$0.20)E-1	\\
\tiny	050404	&\tiny 17:27:48.500	&\tiny (1.14$\pm$0.03)E+1	&\tiny 2.0E-1	&\tiny (1.03$\pm$0.18)E-1	\\
\tiny	050409	&\tiny 01:18:36.050	&\tiny (1.26$\pm$0.05)E+0	&\tiny 2.0E-2	&\tiny (3.05$\pm$0.30)E-1	\\
\tiny	050411	&\tiny 21:51:09.500	&\tiny (6.80$\pm$0.69)E+0	&\tiny 2.0E-1	&\tiny -(1.18$\pm$0.80)E-1	\\
\tiny	050412	&\tiny 18:58:45.900	&\tiny (1.94$\pm$0.08)E+1	&\tiny 2.0E-1	&\tiny -(5.03$\pm$0.51)E-1	\\
\tiny	050429	&\tiny 14:09:50.900	&\tiny (1.94$\pm$0.15)E+1	&\tiny 2.0E-1	&\tiny (1.35$\pm$0.63)E-1	\\
\tiny	050430	&\tiny 09:13:09.100	&\tiny (1.42$\pm$0.17)E+1	&\tiny 2.0E-1	&\tiny (0.20$\pm$9.23)E-2	\\
\tiny	050501	&\tiny 08:19:38.825	&\tiny (2.40$\pm$0.19)E+0	&\tiny 1.5E-1	&\tiny (7.81$\pm$4.03)E-2	\\
\tiny	050502	&\tiny 19:56:57.575	&\tiny (1.60$\pm$0.20)E+0	&\tiny 5.0E-2	&\tiny (1.61$\pm$0.97)E-1	\\
\tiny	050509	&\tiny 09:31:26.700	&\tiny (2.00$\pm$0.04)E+1	&\tiny 2.0E-1	&\tiny -(6.35$\pm$1.57)E-2	\\
\tiny	050516	&\tiny 12:58:05.200	&\tiny (7.60$\pm$1.16)E+0	&\tiny 4.0E-1	&\tiny (1.16$\pm$1.11)E-1	\\

\end{tabular}
\end{table}
\end{center}

\begin{center}
\begin{table} \caption{The RHESSI GRB data-set.}
\centering
\begin{tabular}{lrrrr}

\hline
\\[-2ex]
GRB     & peak time         & $T_{90}$                          & $\delta t_{res}$ & hardness ratio         \\[0.3ex]
        & UTC               & [s]                               & [s]              & log $H$                \\[0.3ex]
\hline\hline
\\[-2ex]
\tiny	050525A	&\tiny 00:02:54.450	&\tiny (7.40$\pm$0.19)E+0	&\tiny 1.0E-1	&\tiny -(4.06$\pm$0.25)E-1	\\
\tiny	050525B	&\tiny 00:50:00.500	&\tiny (1.26$\pm$0.03)E+1	&\tiny 2.0E-1	&\tiny (6.42$\pm$1.15)E-2	\\
\tiny	050528	&\tiny 07:05:23.500	&\tiny (1.57$\pm$0.09)E+2	&\tiny 1.0E+0	&\tiny -(4.77$\pm$0.69)E-1	\\
\tiny	050530	&\tiny 04:44:44.900	&\tiny (2.40$\pm$0.34)E+0	&\tiny 2.0E-1	&\tiny (4.69$\pm$1.10)E-1	\\
\tiny	050531	&\tiny 04:27:26.700	&\tiny (3.32$\pm$0.05)E+1	&\tiny 2.0E-1	&\tiny -(8.44$\pm$1.34)E-2	\\
\tiny	050614	&\tiny 12:02:00.500	&\tiny (3.10$\pm$0.45)E+1	&\tiny 1.0E+0	&\tiny -(1.73$\pm$1.20)E-1	\\
\tiny	050701	&\tiny 14:22:00.950	&\tiny (1.00$\pm$0.08)E+1	&\tiny 1.0E-1	&\tiny (2.54$\pm$0.69)E-1	\\
\tiny	050702	&\tiny 07:47:44.160	&\tiny (1.12$\pm$0.16)E+0	&\tiny 8.0E-2	&\tiny (2.01$\pm$1.01)E-1	\\
\tiny	050703	&\tiny 05:31:50.825	&\tiny (8.65$\pm$0.44)E+0	&\tiny 5.0E-2	&\tiny -(2.24$\pm$0.47)E-1	\\
\tiny	050706	&\tiny 17:11:37.875	&\tiny (6.15$\pm$0.42)E+0	&\tiny 1.5E-1	&\tiny -(0.07$\pm$5.20)E-2	\\
\tiny	050713A	&\tiny 04:29:11.750	&\tiny (1.86$\pm$0.07)E+1	&\tiny 3.0E-1	&\tiny -(9.16$\pm$2.81)E-2	\\
\tiny	050713B	&\tiny 12:07:28.750	&\tiny (6.05$\pm$0.39)E+1	&\tiny 5.0E-1	&\tiny -(0.35$\pm$5.23)E-2	\\
\tiny	050715	&\tiny 01:15:49.250	&\tiny (1.14$\pm$0.03)E+1	&\tiny 1.0E-1	&\tiny -(2.32$\pm$0.20)E-1	\\
\tiny	050717	&\tiny 10:30:55.100	&\tiny (9.60$\pm$0.32)E+0	&\tiny 2.0E-1	&\tiny (2.36$\pm$0.23)E-1	\\
\tiny	050726	&\tiny 20:22:19.800	&\tiny (1.60$\pm$0.06)E+1	&\tiny 4.0E-1	&\tiny -(5.78$\pm$2.30)E-2	\\
\tiny	050729	&\tiny 01:09:41.120	&\tiny (4.40$\pm$0.47)E+0	&\tiny 8.0E-2	&\tiny -(8.49$\pm$8.47)E-2	\\
\tiny	050802	&\tiny 10:08:02.850	&\tiny (2.94$\pm$0.31)E+1	&\tiny 3.0E-1	&\tiny -(1.76$\pm$0.90)E-1	\\
\tiny	050805	&\tiny 13:29:47.625	&\tiny (1.05$\pm$0.08)E+0	&\tiny 5.0E-2	&\tiny (2.99$\pm$0.53)E-1	\\
\tiny	050809	&\tiny 20:15:26.720	&\tiny (2.40$\pm$0.14)E+0	&\tiny 8.0E-2	&\tiny -(7.56$\pm$3.92)E-2	\\
\tiny	050813	&\tiny 21:13:43.900	&\tiny (8.00$\pm$0.87)E+0	&\tiny 2.0E-1	&\tiny (3.68$\pm$0.99)E-1	\\
\tiny	050814	&\tiny 04:35:19.451	&\tiny (1.24$\pm$0.08)E-1	&\tiny 2.0E-3	&\tiny (6.01$\pm$0.64)E-1	\\
\tiny	050817	&\tiny 10:43:18.900	&\tiny (2.12$\pm$0.20)E+1	&\tiny 2.0E-1	&\tiny (3.69$\pm$0.86)E-1	\\
\tiny	050820	&\tiny 23:50:36.050	&\tiny (6.00$\pm$0.58)E+0	&\tiny 3.0E-1	&\tiny -(3.57$\pm$0.87)E-1	\\
\tiny	050824	&\tiny 11:57:42.535	&\tiny (2.50$\pm$0.14)E-1	&\tiny 1.0E-2	&\tiny (3.31$\pm$0.35)E-1	\\
\tiny	050825	&\tiny 03:34:27.850	&\tiny (8.00$\pm$0.44)E-1	&\tiny 2.0E-2	&\tiny (1.62$\pm$0.42)E-1	\\
\tiny	050902	&\tiny 12:24:30.650	&\tiny (1.16$\pm$0.10)E+1	&\tiny 1.0E-1	&\tiny -(5.97$\pm$1.18)E-1	\\
\tiny	050923	&\tiny 01:37:44.850	&\tiny (9.60$\pm$0.56)E+0	&\tiny 1.0E-1	&\tiny (6.55$\pm$4.65)E-2	\\
\tiny	051009	&\tiny 10:49:02.750	&\tiny (8.50$\pm$0.52)E+1	&\tiny 5.0E-1	&\tiny -(2.22$\pm$0.57)E-1	\\
\tiny	051012	&\tiny 12:00:11.925	&\tiny (2.55$\pm$0.08)E+0	&\tiny 5.0E-2	&\tiny (1.14$\pm$1.98)E-2	\\
\tiny	051021	&\tiny 14:01:14.625	&\tiny (5.10$\pm$0.10)E+1	&\tiny 2.5E-1	&\tiny -(2.26$\pm$0.17)E-1	\\
\tiny	051031	&\tiny 22:01:05.750	&\tiny (4.95$\pm$0.28)E+1	&\tiny 1.5E+0	&\tiny (4.49$\pm$3.88)E-2	\\
\tiny	051101	&\tiny 01:13:12.625	&\tiny (1.10$\pm$0.14)E+1	&\tiny 1.5E-1	&\tiny (9.19$\pm$9.99)E-2	\\
\tiny	051103	&\tiny 09:25:42.192	&\tiny (1.40$\pm$0.05)E-1	&\tiny 5.0E-3	&\tiny (4.68$\pm$0.15)E-1	\\
\tiny	051109	&\tiny 16:42:00.750	&\tiny (2.55$\pm$0.16)E+1	&\tiny 5.0E-1	&\tiny -(2.15$\pm$0.58)E-1	\\
\tiny	051111	&\tiny 05:59:40.100	&\tiny (2.38$\pm$0.15)E+1	&\tiny 2.0E-1	&\tiny (1.10$\pm$0.53)E-1	\\
\tiny	051117	&\tiny 12:34:25.300	&\tiny (3.32$\pm$0.13)E+1	&\tiny 2.0E-1	&\tiny (1.62$\pm$0.34)E-1	\\
\tiny	051119	&\tiny 13:10:59.900	&\tiny (3.78$\pm$0.31)E+1	&\tiny 2.0E-1	&\tiny -(1.74$\pm$0.73)E-1	\\
\tiny	051124A	&\tiny 08:16:59.450	&\tiny (1.92$\pm$0.15)E+1	&\tiny 1.0E-1	&\tiny (2.31$\pm$0.65)E-1	\\
\tiny	051124B	&\tiny 14:20:11.100	&\tiny (3.68$\pm$0.19)E+1	&\tiny 2.0E-1	&\tiny -(2.16$\pm$0.48)E-1	\\
\tiny	051201A	&\tiny 18:31:37.250	&\tiny (2.25$\pm$0.29)E+1	&\tiny 1.5E+0	&\tiny -(0.50$\pm$8.84)E-2	\\
\tiny	051201B	&\tiny 22:35:30.350	&\tiny (1.84$\pm$0.13)E+1	&\tiny 1.0E-1	&\tiny (1.29$\pm$0.59)E-1	\\
\tiny	051207	&\tiny 19:04:09.350	&\tiny (5.73$\pm$0.17)E+1	&\tiny 3.0E-1	&\tiny -(1.75$\pm$0.27)E-1	\\
\tiny	051211	&\tiny 05:28:13.250	&\tiny (2.94$\pm$0.09)E+1	&\tiny 7.0E-1	&\tiny (1.24$\pm$0.17)E-1	\\
\tiny	051217	&\tiny 09:54:08.650	&\tiny (1.80$\pm$0.27)E+1	&\tiny 3.0E-1	&\tiny -(0.28$\pm$1.14)E-1	\\
\tiny	051220	&\tiny 13:04:17.525	&\tiny (1.29$\pm$0.01)E+1	&\tiny 5.0E-2	&\tiny (2.15$\pm$0.04)E-1	\\
\tiny	051220	&\tiny 21:34:36.750	&\tiny (2.80$\pm$0.46)E+1	&\tiny 5.0E-1	&\tiny -(5.31$\pm$2.02)E-1	\\
\tiny	051221	&\tiny 01:51:15.975	&\tiny (2.80$\pm$0.17)E-1	&\tiny 1.0E-2	&\tiny (1.65$\pm$0.41)E-1	\\
\tiny	051222	&\tiny 15:07:35.600	&\tiny (2.44$\pm$0.40)E+1	&\tiny 4.0E-1	&\tiny -(1.73$\pm$1.35)E-1	\\
\tiny	060101	&\tiny 00:34:11.800	&\tiny (1.96$\pm$0.09)E+1	&\tiny 4.0E-1	&\tiny (6.61$\pm$3.35)E-2	\\
\tiny	060110	&\tiny 08:01:18.900	&\tiny (1.20$\pm$0.16)E+1	&\tiny 2.0E-1	&\tiny (1.03$\pm$1.00)E-1	\\
\tiny	060111	&\tiny 08:49:00.250	&\tiny (6.90$\pm$0.25)E+1	&\tiny 1.5E+0	&\tiny (1.61$\pm$0.24)E-1	\\
\tiny	060117	&\tiny 06:50:13.850	&\tiny (1.59$\pm$0.10)E+1	&\tiny 3.0E-1	&\tiny -(3.13$\pm$0.63)E-1	\\
\tiny	060121A	&\tiny 04:12:56.500	&\tiny (4.40$\pm$0.63)E+1	&\tiny 1.0E+0	&\tiny (0.17$\pm$1.08)E-1	\\
\tiny	060121B	&\tiny 22:24:56.755	&\tiny (2.38$\pm$0.12)E+0	&\tiny 7.0E-2	&\tiny -(1.05$\pm$0.36)E-1	\\
\tiny	060123	&\tiny 05:05:24.900	&\tiny (1.10$\pm$0.03)E+1	&\tiny 2.0E-1	&\tiny -(1.31$\pm$0.17)E-1	\\
\tiny	060124	&\tiny 16:04:22.400	&\tiny (1.76$\pm$0.13)E+1	&\tiny 4.0E-1	&\tiny (4.99$\pm$5.58)E-2	\\
\tiny	060130	&\tiny 13:48:31.100	&\tiny (1.52$\pm$0.08)E+1	&\tiny 2.0E-1	&\tiny (2.98$\pm$4.17)E-2	\\
\tiny	060203	&\tiny 07:28:58.535	&\tiny (5.40$\pm$0.45)E-1	&\tiny 1.0E-2	&\tiny (1.43$\pm$0.66)E-1	\\
\tiny	060217	&\tiny 09:47:43.325	&\tiny (1.91$\pm$0.08)E+1	&\tiny 1.5E-1	&\tiny -(7.34$\pm$3.62)E-2	\\
\tiny	060224	&\tiny 02:31:11.500	&\tiny (1.00$\pm$0.06)E+1	&\tiny 2.0E-1	&\tiny -(8.14$\pm$4.54)E-2	\\
\tiny	060228	&\tiny 03:17:33.850	&\tiny (3.27$\pm$0.21)E+1	&\tiny 3.0E-1	&\tiny -(3.88$\pm$5.28)E-2	\\
\tiny	060303	&\tiny 22:42:47.525	&\tiny (5.00$\pm$0.53)E-1	&\tiny 5.0E-2	&\tiny (2.56$\pm$0.29)E-1	\\
\tiny	060306	&\tiny 15:22:38.485	&\tiny (9.20$\pm$0.11)E-1	&\tiny 1.0E-2	&\tiny -(1.07$\pm$0.04)E-1	\\
\tiny	060309	&\tiny 14:38:54.250	&\tiny (2.95$\pm$0.30)E+1	&\tiny 5.0E-1	&\tiny (1.32$\pm$0.81)E-1	\\
\tiny	060312A	&\tiny 06:17:21.115	&\tiny (2.40$\pm$0.47)E-1	&\tiny 3.0E-2	&\tiny (2.48$\pm$1.16)E-1	\\
\tiny	060312B	&\tiny 16:44:53.400	&\tiny (6.80$\pm$1.18)E+0	&\tiny 4.0E-1	&\tiny -(2.32$\pm$1.40)E-1	\\

\end{tabular}
\end{table}
\end{center}

\begin{center}
\begin{table} \caption{The RHESSI GRB data-set.}
\centering
\begin{tabular}{lrrrr}

\hline
\\[-2ex]
GRB     & peak time         & $T_{90}$                          & $\delta t_{res}$ & hardness ratio         \\[0.3ex]
        & UTC               & [s]                               & [s]              & log $H$                \\[0.3ex]
\hline\hline
\\[-2ex]
\tiny	060313	&\tiny 20:11:32.900	&\tiny (3.60$\pm$0.51)E+0	&\tiny 2.0E-1	&\tiny (1.52$\pm$1.03)E-1	\\
\tiny	060323	&\tiny 07:04:30.100	&\tiny (1.76$\pm$0.04)E+1	&\tiny 2.0E-1	&\tiny (3.37$\pm$1.52)E-2	\\
\tiny	060325	&\tiny 12:02:20.350	&\tiny (1.08$\pm$0.02)E+1	&\tiny 1.0E-1	&\tiny (0.70$\pm$1.11)E-2	\\
\tiny	060401	&\tiny 05:40:18.750	&\tiny (6.30$\pm$0.17)E+0	&\tiny 1.0E-1	&\tiny (9.21$\pm$1.86)E-2	\\
\tiny	060408	&\tiny 13:11:39.150	&\tiny (6.90$\pm$0.95)E+0	&\tiny 3.0E-1	&\tiny -(0.06$\pm$1.01)E-1	\\
\tiny	060415	&\tiny 05:31:00.050	&\tiny (1.32$\pm$0.20)E+1	&\tiny 3.0E-1	&\tiny (1.80$\pm$1.14)E-1	\\
\tiny	060418	&\tiny 03:06:35.800	&\tiny (4.08$\pm$0.21)E+1	&\tiny 4.0E-1	&\tiny -(1.23$\pm$0.43)E-1	\\
\tiny	060421A	&\tiny 11:03:49.000	&\tiny (3.12$\pm$0.11)E+1	&\tiny 4.0E-1	&\tiny -(8.58$\pm$2.76)E-2	\\
\tiny	060421B	&\tiny 20:36:38.200	&\tiny (2.16$\pm$0.19)E+1	&\tiny 4.0E-1	&\tiny -(1.39$\pm$0.74)E-1	\\
\tiny	060425	&\tiny 16:57:38.705	&\tiny (1.40$\pm$0.12)E-1	&\tiny 1.0E-2	&\tiny (2.59$\pm$0.42)E-1	\\
\tiny	060428	&\tiny 02:30:41.750	&\tiny (1.35$\pm$0.23)E+1	&\tiny 5.0E-1	&\tiny -(2.45$\pm$1.44)E-1	\\
\tiny	060429	&\tiny 12:19:51.250	&\tiny (2.00$\pm$0.21)E-1	&\tiny 2.0E-2	&\tiny (2.89$\pm$0.33)E-1	\\
\tiny	060505	&\tiny 23:32:01.050	&\tiny (9.60$\pm$0.46)E+0	&\tiny 3.0E-1	&\tiny (2.15$\pm$0.30)E-1	\\
\tiny	060528	&\tiny 22:53:05.750	&\tiny (7.70$\pm$0.20)E+1	&\tiny 5.0E-1	&\tiny (2.41$\pm$0.21)E-1	\\
\tiny	060530	&\tiny 19:19:11.300	&\tiny (4.00$\pm$0.35)E+0	&\tiny 2.0E-1	&\tiny -(1.16$\pm$0.63)E-1	\\
\tiny	060610	&\tiny 11:22:24.070	&\tiny (6.00$\pm$0.33)E-1	&\tiny 2.0E-2	&\tiny (2.48$\pm$0.38)E-1	\\
\tiny	060614	&\tiny 12:43:48.250	&\tiny (5.25$\pm$0.45)E+1	&\tiny 1.5E+0	&\tiny -(1.78$\pm$0.72)E-1	\\
\tiny	060622	&\tiny 17:19:48.750	&\tiny (2.35$\pm$0.09)E+1	&\tiny 5.0E-1	&\tiny (1.12$\pm$0.27)E-1	\\
\tiny	060624	&\tiny 13:46:56.255	&\tiny (2.55$\pm$0.07)E+0	&\tiny 3.0E-2	&\tiny -(4.55$\pm$0.29)E-1	\\
\tiny	060625	&\tiny 07:33:27.100	&\tiny (4.40$\pm$0.37)E+0	&\tiny 2.0E-1	&\tiny (2.21$\pm$0.58)E-1	\\
\tiny	060630	&\tiny 00:06:41.250	&\tiny (4.10$\pm$0.19)E+1	&\tiny 5.0E-1	&\tiny -(9.96$\pm$4.00)E-2	\\
\tiny	060708	&\tiny 04:30:38.485	&\tiny (1.14$\pm$0.08)E-1	&\tiny 6.0E-3	&\tiny (1.57$\pm$0.36)E-1	\\
\tiny	060729	&\tiny 04:07:38.600	&\tiny (5.20$\pm$0.53)E+0	&\tiny 2.0E-1	&\tiny -(1.16$\pm$0.79)E-1	\\
\tiny	060805	&\tiny 14:27:17.450	&\tiny (5.10$\pm$0.07)E+0	&\tiny 6.0E-2	&\tiny (6.98$\pm$0.69)E-2	\\
\tiny	060811	&\tiny 16:56:43.950	&\tiny (7.29$\pm$0.27)E+1	&\tiny 3.0E-1	&\tiny -(1.96$\pm$0.35)E-1	\\
\tiny	060819	&\tiny 18:28:20.700	&\tiny (2.40$\pm$0.11)E+1	&\tiny 6.0E-1	&\tiny (0.84$\pm$3.26)E-2	\\
\tiny	060823	&\tiny 08:05:33.750	&\tiny (1.00$\pm$0.18)E+0	&\tiny 1.0E-1	&\tiny (3.23$\pm$1.26)E-1	\\
\tiny	060919	&\tiny 21:52:12.750	&\tiny (2.95$\pm$0.37)E+1	&\tiny 5.0E-1	&\tiny (1.20$\pm$0.95)E-1	\\
\tiny	060920	&\tiny 15:32:38.700	&\tiny (2.18$\pm$0.03)E+1	&\tiny 2.0E-1	&\tiny (3.59$\pm$1.01)E-2	\\
\tiny	060925	&\tiny 20:14:35.375	&\tiny (1.65$\pm$0.08)E+1	&\tiny 2.5E-1	&\tiny -(1.00$\pm$0.38)E-1	\\
\tiny	060928	&\tiny 01:20:24.150	&\tiny (2.03$\pm$0.03)E+2	&\tiny 3.0E-1	&\tiny -(1.28$\pm$1.19)E-2	\\
\tiny	061005	&\tiny 13:38:01.800	&\tiny (4.24$\pm$0.12)E+1	&\tiny 4.0E-1	&\tiny -(2.05$\pm$0.25)E-1	\\
\tiny	061006A	&\tiny 08:43:39.225	&\tiny (1.65$\pm$0.10)E+0	&\tiny 5.0E-2	&\tiny (1.53$\pm$0.43)E-1	\\
\tiny	061006B	&\tiny 16:45:27.875	&\tiny (4.00$\pm$0.53)E-1	&\tiny 5.0E-2	&\tiny (3.23$\pm$0.39)E-1	\\
\tiny	061007	&\tiny 10:08:54.150	&\tiny (6.06$\pm$0.07)E+1	&\tiny 3.0E-1	&\tiny (1.20$\pm$0.08)E-1	\\
\tiny	061012	&\tiny 11:51:57.850	&\tiny (9.30$\pm$0.48)E+0	&\tiny 1.0E-1	&\tiny (2.56$\pm$0.44)E-1	\\
\tiny	061013	&\tiny 18:06:28.200	&\tiny (4.00$\pm$0.22)E+1	&\tiny 8.0E-1	&\tiny -(1.54$\pm$0.45)E-1	\\
\tiny	061014	&\tiny 06:17:02.375	&\tiny (2.00$\pm$0.55)E-1	&\tiny 5.0E-2	&\tiny (2.56$\pm$0.94)E-1	\\
\tiny	061022	&\tiny 12:23:42.850	&\tiny (2.19$\pm$0.29)E+1	&\tiny 3.0E-1	&\tiny -(0.44$\pm$1.02)E-1	\\
\tiny	061031	&\tiny 12:19:51.500	&\tiny (3.30$\pm$0.12)E+1	&\tiny 2.0E-1	&\tiny (7.92$\pm$3.06)E-2	\\
\tiny	061101	&\tiny 21:26:51.550	&\tiny (1.92$\pm$0.30)E+1	&\tiny 3.0E-1	&\tiny -(2.41$\pm$1.40)E-1	\\
\tiny	061108	&\tiny 01:09:54.850	&\tiny (3.75$\pm$0.09)E+1	&\tiny 3.0E-1	&\tiny -(8.68$\pm$1.90)E-2	\\
\tiny	061113	&\tiny 13:43:36.050	&\tiny (1.82$\pm$0.03)E+1	&\tiny 1.0E-1	&\tiny (1.76$\pm$0.13)E-1	\\
\tiny	061117	&\tiny 06:00:11.250	&\tiny (2.15$\pm$0.27)E+1	&\tiny 5.0E-1	&\tiny (9.86$\pm$9.49)E-2	\\
\tiny	061121	&\tiny 15:23:44.275	&\tiny (1.46$\pm$0.03)E+1	&\tiny 1.5E-1	&\tiny (1.27$\pm$0.17)E-1	\\
\tiny	061123	&\tiny 16:33:28.650	&\tiny (5.70$\pm$0.32)E+0	&\tiny 1.0E-1	&\tiny (1.23$\pm$0.43)E-1	\\
\tiny	061126	&\tiny 08:48:03.150	&\tiny (1.65$\pm$0.07)E+1	&\tiny 3.0E-1	&\tiny (2.49$\pm$0.31)E-1	\\
\tiny	061128	&\tiny 20:01:11.805	&\tiny (3.00$\pm$0.31)E-1	&\tiny 3.0E-2	&\tiny (3.50$\pm$0.27)E-1	\\
\tiny	061205	&\tiny 05:22:15.450	&\tiny (7.50$\pm$0.91)E+0	&\tiny 3.0E-1	&\tiny -(1.22$\pm$0.96)E-1	\\
\tiny	061212	&\tiny 05:31:30.970	&\tiny (1.90$\pm$0.01)E+1	&\tiny 6.0E-2	&\tiny (2.73$\pm$0.06)E-1	\\
\tiny	061222	&\tiny 03:30:19.300	&\tiny (1.16$\pm$0.05)E+1	&\tiny 2.0E-1	&\tiny (1.44$\pm$0.31)E-1	\\
\tiny	061229	&\tiny 22:25:44.250	&\tiny (7.95$\pm$0.97)E+1	&\tiny 5.0E-1	&\tiny (1.69$\pm$0.93)E-1	\\
\tiny	061230	&\tiny 23:09:31.000	&\tiny (2.56$\pm$0.19)E+1	&\tiny 8.0E-1	&\tiny -(1.63$\pm$0.60)E-1	\\
\tiny	070113	&\tiny 11:56:23.815	&\tiny (2.70$\pm$0.48)E-1	&\tiny 3.0E-2	&\tiny (1.65$\pm$1.04)E-1	\\
\tiny	070116	&\tiny 14:32:16.125	&\tiny (1.65$\pm$0.11)E+1	&\tiny 2.5E-1	&\tiny (3.79$\pm$5.24)E-2	\\
\tiny	070120	&\tiny 10:48:36.250	&\tiny (1.85$\pm$0.27)E+1	&\tiny 5.0E-1	&\tiny (1.94$\pm$1.07)E-1	\\
\tiny	070121	&\tiny 10:12:17.000	&\tiny (8.80$\pm$1.60)E+0	&\tiny 8.0E-1	&\tiny (0.11$\pm$1.19)E-1	\\
\tiny	070125	&\tiny 07:21:27.250	&\tiny (5.67$\pm$0.08)E+1	&\tiny 3.0E-1	&\tiny (5.03$\pm$1.12)E-2	\\
\tiny	070214	&\tiny 22:39:20.850	&\tiny (1.77$\pm$0.27)E+1	&\tiny 3.0E-1	&\tiny (1.00$\pm$1.11)E-1	\\
\tiny	070220	&\tiny 04:44:45.300	&\tiny (2.14$\pm$0.11)E+1	&\tiny 2.0E-1	&\tiny (4.86$\pm$4.03)E-2	\\
\tiny	070221	&\tiny 21:06:46.500	&\tiny (1.02$\pm$0.13)E+1	&\tiny 2.0E-1	&\tiny -(2.28$\pm$9.64)E-2	\\
\tiny	070307	&\tiny 21:15:43.250	&\tiny (5.25$\pm$0.27)E+1	&\tiny 5.0E-1	&\tiny -(5.80$\pm$4.35)E-2	\\
\tiny	070402	&\tiny 15:48:39.475	&\tiny (8.85$\pm$0.56)E+0	&\tiny 1.5E-1	&\tiny (4.96$\pm$4.92)E-2	\\
\tiny	070420	&\tiny 06:18:18.400	&\tiny (6.16$\pm$0.36)E+1	&\tiny 8.0E-1	&\tiny -(1.63$\pm$0.51)E-1	\\
\tiny	070508	&\tiny 04:18:25.050	&\tiny (1.31$\pm$0.04)E+1	&\tiny 1.0E-1	&\tiny -(1.71$\pm$2.74)E-2	\\
\tiny	070516	&\tiny 20:41:24.725	&\tiny (3.50$\pm$0.56)E-1	&\tiny 5.0E-2	&\tiny (5.01$\pm$0.73)E-1	\\

\end{tabular}
\end{table}
\end{center}

\begin{center}
\begin{table} \caption{The RHESSI GRB data-set.}
\label{data-last}
\centering
\begin{tabular}{lrrrr}

\hline
\\[-2ex]
GRB     & peak time         & $T_{90}$                          & $\delta t_{res}$ & hardness ratio         \\[0.3ex]
        & UTC               & [s]                               & [s]              & log $H$                \\[0.3ex]
\hline\hline
\\[-2ex]
\tiny	070531	&\tiny 11:45:43.500	&\tiny (2.60$\pm$0.34)E+1	&\tiny 1.0E+0	&\tiny -(1.08$\pm$1.03)E-1	\\
\tiny	070614	&\tiny 05:05:09.425	&\tiny (1.50$\pm$0.51)E-1	&\tiny 5.0E-2	&\tiny (3.35$\pm$0.64)E-1	\\
\tiny	070622	&\tiny 02:25:17.850	&\tiny (1.38$\pm$0.03)E+1	&\tiny 1.0E-1	&\tiny (1.23$\pm$0.18)E-1	\\
\tiny	070626	&\tiny 04:08:44.500	&\tiny (1.43$\pm$0.02)E+2	&\tiny 1.0E+0	&\tiny (5.00$\pm$1.07)E-2	\\
\tiny	070710	&\tiny 08:22:07.850	&\tiny (3.90$\pm$0.58)E+0	&\tiny 3.0E-1	&\tiny (0.94$\pm$9.91)E-2	\\
\tiny	070717	&\tiny 21:50:38.750	&\tiny (1.75$\pm$0.20)E+1	&\tiny 5.0E-1	&\tiny (2.50$\pm$0.87)E-1	\\
\tiny	070722	&\tiny 06:00:31.500	&\tiny (7.20$\pm$0.96)E+0	&\tiny 2.0E-1	&\tiny (1.02$\pm$0.99)E-1	\\
\tiny	070724	&\tiny 23:25:46.500	&\tiny (2.40$\pm$0.25)E+1	&\tiny 1.0E+0	&\tiny -(2.08$\pm$0.89)E-1	\\
\tiny	070802	&\tiny 06:16:19.390	&\tiny (3.12$\pm$0.25)E+0	&\tiny 6.0E-2	&\tiny (4.80$\pm$0.78)E-1	\\
\tiny	070817	&\tiny 14:43:42.250	&\tiny (8.80$\pm$0.71)E+1	&\tiny 5.0E-1	&\tiny -(9.18$\pm$6.84)E-2	\\
\tiny	070819	&\tiny 10:17:04.750	&\tiny (3.30$\pm$0.26)E+1	&\tiny 5.0E-1	&\tiny (3.83$\pm$6.10)E-2	\\
\tiny	070821	&\tiny 12:51:31.750	&\tiny (6.75$\pm$0.21)E+1	&\tiny 5.0E-1	&\tiny (6.03$\pm$2.47)E-2	\\
\tiny	070824	&\tiny 20:50:10.425	&\tiny (1.30$\pm$0.07)E+0	&\tiny 5.0E-2	&\tiny (1.01$\pm$0.29)E-1	\\
\tiny	070825	&\tiny 01:55:54.250	&\tiny (3.42$\pm$0.09)E+1	&\tiny 3.0E-1	&\tiny (1.44$\pm$0.19)E-1	\\
\tiny	070917	&\tiny 09:40:31.250	&\tiny (2.31$\pm$0.24)E+1	&\tiny 3.0E-1	&\tiny (3.20$\pm$0.85)E-1	\\
\tiny	071013	&\tiny 08:53:39.475	&\tiny (3.65$\pm$0.43)E+0	&\tiny 5.0E-2	&\tiny (3.27$\pm$8.94)E-2	\\
\tiny	071014	&\tiny 03:19:52.450	&\tiny (7.80$\pm$0.37)E+0	&\tiny 1.0E-1	&\tiny (4.13$\pm$3.71)E-2	\\
\tiny	071030	&\tiny 08:52:41.900	&\tiny (6.00$\pm$0.91)E+0	&\tiny 2.0E-1	&\tiny (2.20$\pm$1.14)E-1	\\
\tiny	071104	&\tiny 11:41:09.525	&\tiny (1.47$\pm$0.13)E+1	&\tiny 1.5E-1	&\tiny -(6.66$\pm$7.12)E-2	\\
\tiny	071204	&\tiny 05:58:29.475	&\tiny (3.00$\pm$0.56)E-1	&\tiny 5.0E-2	&\tiny (3.46$\pm$0.72)E-1	\\
\tiny	071217	&\tiny 17:03:27.950	&\tiny (8.30$\pm$0.48)E+0	&\tiny 1.0E-1	&\tiny -(9.40$\pm$4.89)E-2	\\
\tiny	080114	&\tiny 16:10:22.300	&\tiny (7.34$\pm$0.07)E+1	&\tiny 2.0E-1	&\tiny (3.60$\pm$0.09)E-1	\\
\tiny	080202	&\tiny 13:04:37.250	&\tiny (3.06$\pm$0.39)E+1	&\tiny 3.0E-1	&\tiny (0.15$\pm$9.87)E-2	\\
\tiny	080204	&\tiny 13:56:34.760	&\tiny (4.88$\pm$0.23)E+0	&\tiny 8.0E-2	&\tiny (4.53$\pm$0.42)E-1	\\
\tiny	080211	&\tiny 07:23:46.250	&\tiny (2.80$\pm$0.08)E+1	&\tiny 5.0E-1	&\tiny (2.79$\pm$0.19)E-1	\\
\tiny	080218	&\tiny 05:57:28.375	&\tiny (1.95$\pm$0.15)E+1	&\tiny 2.5E-1	&\tiny (1.18$\pm$0.61)E-1	\\
\tiny	080224	&\tiny 16:58:51.050	&\tiny (5.40$\pm$0.29)E+0	&\tiny 1.0E-1	&\tiny (2.22$\pm$0.41)E-1	\\
\tiny	080318	&\tiny 08:31:45.050	&\tiny (1.47$\pm$0.16)E+1	&\tiny 3.0E-1	&\tiny (1.51$\pm$0.79)E-1	\\
\tiny	080319	&\tiny 12:25:56.900	&\tiny (1.20$\pm$0.06)E+1	&\tiny 2.0E-1	&\tiny (3.29$\pm$0.42)E-1	\\
\tiny	080320	&\tiny 11:52:02.625	&\tiny (3.05$\pm$0.08)E+1	&\tiny 2.5E-1	&\tiny (1.96$\pm$0.22)E-1	\\
\tiny	080328	&\tiny 08:03:14.500	&\tiny (8.60$\pm$0.62)E+1	&\tiny 1.0E+0	&\tiny (3.03$\pm$0.58)E-1	\\
\tiny	080330	&\tiny 11:04:33.450	&\tiny (3.36$\pm$0.09)E+1	&\tiny 3.0E-1	&\tiny (2.54$\pm$0.20)E-1	\\
\tiny	080408	&\tiny 03:36:23.050	&\tiny (1.40$\pm$0.11)E+0	&\tiny 1.0E-1	&\tiny (5.29$\pm$0.37)E-1	\\
\tiny	080413	&\tiny 08:51:12.250	&\tiny (7.60$\pm$1.20)E+0	&\tiny 1.0E-1	&\tiny (0.55$\pm$1.17)E-1	\\
\tiny	080425	&\tiny 20:21:47.775	&\tiny (2.28$\pm$0.32)E+1	&\tiny 1.5E-1	&\tiny (2.58$\pm$1.05)E-1	\\

\end{tabular}
\end{table}
\end{center}

\end{document}